\documentclass{template}
\usepackage{graphicx,parskip,appendix,float}
\usepackage[ruled] {algorithm2e}
\usepackage{url,amsmath,amssymb,fancybox,listings,pdfpages,multicol,multirow,datetime,rotating, booktabs, geometry, enumitem, xcolor, soul, tikz,adjustbox}
\usepackage{subcaption}
\usepackage[pagebackref=false,pdffitwindow=true]{hyperref}
\usepackage{csquotes}

\usepackage{caption}
\usepackage[style=numeric,backend=biber]{biblatex}% \usepackage[nounderscore]{syntax}
\tablespagefalse
\listingspagetrue
\figurespagetrue

\setlist[itemize]{noitemsep, label={\raisebox{2pt}{\tiny\textbullet}}}

\usepackage{changepage} % allows stuff to exceed margin, adjustwidth
 
\usepackage[nomain,section,savewrites,nopostdot,nogroupskip,acronym,style=super,nonumberlist,xindy]{glossaries-extra}
\makeglossaries
\setabbreviationstyle[acronym]{long-short}
\newacronym{mit}{MIT}{Massachusetts Institute of Technology}

\newacronym{au}{AU}{Aarhus University}

\newacronym{vpn}{VPN}{virtual private network}

\newacronym{dry}{DRY}{don't repeat yourself}

\newacronym{kiss}{KISS}{keep it simple, stupid}

\newacronym{solid}{SOLID}{single responsibility, open-closed, Liskov substitution, interface segregation, dependency inversion}

\newacronym{api}{API}{application programming interface}

\newacronym{dsl}{DSL}{domain-specific language}

\newacronym{csp}{CSP}{communicating sequential processes}

\newacronym{sat}{SAT}{satisfiability}

\newacronym{ltl}{LTL}{linear temporal logic}

\newacronym{ast}{AST}{abstract syntax tree}

\newacronym{jvm}{JVM}{Java Virtual Machine}

\newacronym{ml}{ML}{Meta Language}

\newacronym{js}{JS}{JavaScript}

\newacronym{ascii}{ASCII}{American Standard Code for Information Interchange}

\newacronym{nfa}{NFA}{nondeterministic finite automaton}

\newacronym{dfa}{DFA}{deterministic finite automaton}

\newacronym{php}{PHP}{Hypertext Preprocessor}

\newacronym{yacc}{Yacc}{Yet Another Compiler Compiler}

\newacronym{bnf}{BNF}{Backus–Naur form}

\newacronym{llvm}{LLVM}{Low Level Virtual Machine}

\newacronym{moscow}{MoSCoW}{must have, should have, could have and will not have}

\newacronym{json}{JSON}{JavaScript Object Notation}

\newacronym{html}{HTML}{Hyper Text Markup Language}

\newacronym{mean}{MEAN}{MongoDB-Express-Angular-Node.js}

\newacronym{csail}{CSAIL}{Computer Science and Artificial Intelligence Laboratory}

\newacronym{gpt}{GPT}{Generative Pre-trained Transformer}

\newacronym{llm}{LLM}{large language model}

\newacronym{git}{Git}{global information tracker}

\newacronym{ir}{IR}{intermediate representation}

\newacronym{ide}{IDE}{integrated development environment}

\newacronym{lsp}{LSP}{language server protocol}

\newacronym{xml}{XML}{Extensible Markup Language}

\newacronym{vdm}{VDM}{Vienna Development Method}

\newacronym{vista}{VISTA}{Visiting Student Association}

\newacronym{mde}{MDE}{model-driven engineering}

\newacronym{rest}{REST}{REpresentational State Transfer}

\newacronym{crud}{CRUD}{create-read-update-delete}
\glssetcategoryattribute{acronym}{glossdesc}{title}
\MFUhyphentrue

\newcommand*{\noaddvspace}{\renewcommand*{\addvspace}[1]{}}
\addtocontents{lof}{\protect\noaddvspace}

\usepackage[ligature, inference]{semantic}

\usepackage{grammar} % parsar grammar
\newenvironment{myalign}{%
    \noindent
    \begin{tabular}{@{\hspace{2.5em}}l@{\hspace{2.5em}}l}
}{
    \end{tabular}
}

\hypersetup{
    pdftitle    = {Conceptual},
    pdfauthor   = {Nikolaj Kühne Jakobsen},
    pdfsubject  = {Computer Science},
    pdfkeywords = {Compiler, modeling, Alloy, concepts},
    colorlinks  = true, anchorcolor = blue, filecolor = blue, urlcolor = blue,
    linkcolor   = colIdentifier,    
    citecolor   = colIdentifier,    
}

\definecolor{colBackGrnd}{rgb}{1,1,0.8}
\definecolor{colKeys}{rgb}{0,0,1}
\definecolor{colIdentifier}{rgb}{0,0,0}
\definecolor{colComments}{rgb}{0,.5,0}
\definecolor{colString}{rgb}{0,0,1}
\definecolor{colWhite}{rgb}{1,1,1}
\definecolor{light-gray}{gray}{0.85}

\newcommand{\MyHookSign}{\hbox{\ensuremath\hookleftarrow}}

\lstdefinestyle{kw}{
  stringstyle=\color{orange},
  basicstyle=\ttfamily,
  keywordstyle=,
}

\newsavebox{\kwbox} 
\newcommand{\kw}[1]{%
  \sbox{\kwbox}{\lstinline[style=kw]|#1|}% Store the lstinline content
  \colorbox{light-gray}{\usebox{\kwbox}}% Place the content inside a colorbox
}

\lstdefinestyle{listingstyle}{
columns=[c]fixed,
basicstyle=
\fontfamily{pcr}\selectfont\small,% Change the font to allow keywords to be highlighted
keywordstyle=\bfseries, %
upquote=true,
commentstyle=,
breaklines=true,
prebreak={\space\MyHookSign},
backgroundcolor=\color{colBackGrnd},
breakautoindent=true, %
captionpos=b,%
belowskip=-5pt,
columns=flexible,
tabsize=2,
frame=single,
extendedchars=true, %
showspaces=false,
showstringspaces=false,
numbers=left, %
numberstyle=\footnotesize,
showstringspaces=false,
literate=    {\ \ }{{\ }}1
             {~}{$\sim$}1
             {\#}{$\#$}1
             {->}{$\rightarrow$\ }1
             {=>}{$\Rightarrow$}2,
}
\lstdefinestyle{concept}{
style=listingstyle, % base on this style 
morekeywords={concept, purpose, state, actions, principle,app,include,sync, 
and, or, when, can, until, then, no, is, empty, none, in, not, set, one, lone, some, const, string, int}, 
}

\lstdefinestyle{alloy}{
style=listingstyle,
morekeywords={abstract, after, all, always, and, as, assert, before, but, check, disj, else, enabled, event, eventually, exactly, extends, fact, for, fun, historically, iden, iff, implies, in, Int, String, invariant, let, lone, modifies, module, no, none, not, once, one, open, or, pred, releases, run, set, sig, since, some, steps, sum, triggered, univ, until, var
}
}

\begin{document}

\DeclareGraphicsExtensions{.jpg,.png,.gif,.pdf}
%NOTE: When inserting Figures if the extension of the graphic file is not provided LaTeX will automatically search
% for the extensions declared above, in the order declared.

\title{\huge{Establishing tool support for a concept DSL}}

\degreetitle{BSc in Computer Engineering} % Replace with appropriate degree
\rpttype{BSc}    % Replace MSc with BSc for Honours Degree Year projects.

\beforeabstract
\prefacesection{Abstract}
The quality of software products tends to correlate with the quality of the abstractions adopted early in the design process. Acknowledging this tendency has led to the development of various tools and methodologies for modeling systems thoroughly before implementing them. However, creating effective abstract models of domain problems is difficult, especially if the models are also expected to exhibit qualities such as intuitiveness, being seamlessly integrable with other models, or being easily translatable into code.

This thesis describes \textit{Conceptual}, a \gls{dsl} for modeling the behavior of software systems using self-contained and highly reusable units of functionally known as \textit{concepts}. The language's syntax and semantics are formalized based on previous work. Additionally, the thesis proposes a strategy for mapping language constructs from \textit{Conceptual} into the Alloy modeling language. The suggested strategy is then implemented with a simple compiler, allowing developers to access and utilize Alloy's existing analysis tools for program reasoning. 

The utility and expressiveness of \textit{Conceptual} is demonstrated qualitatively through several practical case studies. Using the implemented compiler, a few erroneous specifications are identified in the literature. Moreover, the thesis establishes preliminary tool support in the Visual Studio Code \gls{ide}.

\prefacesection{Acknowledgements}
First and foremost, I wish to express my sincere gratitude to my academic supervisors, who have shown remarkable courage and trust by letting me put their names on this thesis. I am particularly grateful to Peter Gorm Larsen for guiding me through the process, his willingness to proofread my work, and for providing invaluable feedback. Likewise, I am incredibly thankful to Daniel Jackson for hosting me, for his ability to clearly convey technical insights and his deep understanding of software design, and for including me in various of his group's research meetings. Participating in these meetings has been an enriching and informative experience. Additionally, I'm grateful to Nuno Macedo for answering a few technical questions related to Alloy6. 

I also wish to thank the MIT community for being so welcoming and helpful. Thank you for sponsoring my J1 visa and for being such a wonderful research hub. Thank you to Ashlee Andrews and Stella Rupia for their extended help expediting the application process once the hold on the visiting student program was lifted. Additional acknowledgments go to Qi Zhang, Maibritt Hjorth, and Sara Lassen from Aarhus University for helping me with the unusual planning of an undergraduate thesis abroad. Thank you to my colleagues at Aarhus University for consistently keeping me posted regarding important deadlines and happenings at home. Without your updates, I would have missed the application window for the master's program, among other things. 

I would also like to extend my thanks to the foundations, institutions, and organizations that helped fund the trip: \textit{STIBOFONDEN}, \textit{Reinholdt W. Jorck og Hustrus Fond}, \textit{IDA}, \textit{It-vest}, \textit{Jyske Banks almennyttige Fond} and the committee for the Queen's travel grant at Aarhus University. Without your support, this would not have happened!  

Lastly, I want to express my heartfelt gratitude to my friends and family, who have shown great interest in my stay across the Atlantic, kept in touch, or even visited me in the United States. For all of this, thank you!

\afterpreface \afterabstract

% \lstlistoflistings  %NOTE: Will generate a list of Program Listings in the Table of Contents Section
% \listofalgorithms   %NOTE: Will generate a list of Algorithms in the Table of Contents Section

%NOTE: Include the relative reference for each chapter to be included
% dividing the thesis file structure into a number of directories aids the development
% format: directoryName/filename (the .tex extension is not required for the filename)

\chapter{Introduction}\label{chap:Introduction}
\pagenumbering{arabic} \setcounter{page}{1}
\textit{This chapter aims to introduce the subject of the thesis and its significance within the broader field of modeling. First, section~\ref{sec:overview} presents an overview of the software development landscape, depicting the need for new methods or tools. Subsequently, section~\ref{sec:motivation} outlines the research's motivation and highlights the strengths and limitations of a concept-based design approach. Then, section~\ref{sec:goals} defines the study's goals. Following this, section~\ref{sec:reading_guide} aims to assist readers by outlining how different types of information are presented in the thesis. Finally, the chapter concludes with an overview of the thesis structure in section~\ref{sec:structure}.}

\section{Overview}\label{sec:overview}
The second law of thermodynamics implies that entropy tends toward a maximum. Developers may perceive themselves as transcending such physical laws, yet the inexorable increase in disorder appears to manifest within computer systems too \cite{HuntThomas99_pragmatic}. In principle, software engineering can be considered a reversible process, but reverting large-scale systems to previous states can quickly become impractical or even impossible without significant effort. Additionally, the push for market-ready products can eclipse thorough testing and quality assurance. And since "great software today is preferable to perfect software tomorrow," as Andrew Hunt and David Thomas put it in \cite{HuntThomas99_pragmatic}, design flaws may be swiftly patched rather than fundamentally resolved. 

Acknowledging the positive effects of getting the abstractions right early in the design process \cite{DanielJacksonSoftwareAbstractions}, extensive research and discussion about how to create effective designs have taken place. The Gang of Four famously cataloged a set of object-oriented patterns for solving recurring design problems (mostly related to reducing coupling) \cite{GangOfFourPatterns}. Nevertheless, the patterns solve problems that are not necessarily behavioral in nature, thus requiring a level of familiarity to use them. The same thing applies at a higher level with architectural styles and patterns \cite{shaw1996software_architecture_styles}, whereby adhering to a particular style, one can avoid the complexities of ad hoc structuring and achieve certain functional properties. And while design gurus describe various principles for writing effective code \cite{martin2000design_SOLID,kiss2017remarksGiancarlo,HuntThomas99_pragmatic}, often the key tenets are nothing new but rather pertain to old knowledge, echoing the importance of modularity and simplicity \cite{dijkstra1972humble,dijkstra1982role_of_scientific_thought}. Beyond increasing readability, maintainability, and extensibility \cite{Sullivan2001ModularityValue}, writing small components that avoid intelligent tricks makes them easier to reuse. Reusing quality artifacts, in turn, not only saves development time but can also enhance the quality of the software system \cite{Krueger1992Reuse}. 

Today, many popular software applications share a seemingly similar set of features. However, upon inspection, the features are often tailored to the specific platform. For example, favoriting a song on Spotify affects personalized recommendations, but the same feature serves only as a bookmark on Hacker News. Different requirements may necessitate unique implementations, hindering reuse, increasing technical debt \cite{TechnicalDebt_package_changing_metric}, and decreasing modularity \cite{TechnicalDebt_modularity}. Consequently, reusing software across domains is not always effective \cite{TechnicalDebt_StackOverflow}, and even traditional paradigms, though grounded in design patterns and modularity principles, can fall short. 

In his latest book \cite{JacksonConcepts}, Daniel Jackson scrutinizes existing industry-grade software, showcasing that even industry giants such as Google, Apple, and Dropbox are not immune to producing products with suboptimal designs. A key issue of the designs in \cite{JacksonConcepts} appears to be unintuitiveness. Conflicting ideas of what software should do versus what it does create complexity and cause unintended interactions to occur. Despite some designs aligning naturally with the expectations of their users, thus making them easier to develop or use effectively, others often have invisible effects that are hard to anticipate. Daniel proposes an alternative way of thinking about software design using \textit{concepts} \cite{JacksonConcepts}. These \textit{concepts} are structures that give coherent accounts of the consequences of actions in a system, often aligning naturally with mental models of its behavior. Concepts are similar to design patterns in many regards: they are created to solve a particular design problem, they provide a common vocabulary, and they provide basic building blocks for constructing reusable software through abstraction. Unlike patterns, which are often categorized as behavioral, creational, or structural, concepts are purely behavioral and user-facing. However, their most significant quality, which makes concepts a basis for software development, is that they are highly reusable due to strictly enforcing modularity. For example, a restaurant service may be considered an amalgamation of concepts such as the \textit{reservation} concept. That concept's general purpose can be reduced to "efficiently use a pool of limited resources," making it widely applicable in other domains such as computer networking or railway signaling. 

This undergraduate thesis explores the possibility of seamless software modeling via the composition of concepts. The work presented in this thesis encompasses various topics, but the primary focuses are software modeling, compiler, and language design. Specifically, the thesis formalizes a custom \gls{dsl}, \textit{Conceptual}, and implements a compiler from this language to Alloy \cite{DanielJacksonAlloy,DanielJacksonSoftwareAbstractions}. The practical application of the \gls{dsl} and the accompanying compiler is then examined through case studies from \cite{JacksonConcepts} and beyond.

\section{Motivation}\label{sec:motivation}
This thesis is motivated by the premise that many software development errors can be avoided by adopting an alternative approach to design thinking. It explores the potential benefits of reimagining the design process through the underlying concepts that constitute the desired behavior as opposed to conventional model-driven methodologies. 

It appears natural that aligning the design with simple and familiar concepts would help streamline development and eliminate unintended interactions. Especially since these concepts deal with foundational ideas and principles that can be universally applied across various domains and implementations. However, it is undeniable that people can have significantly different mental models of how something works or even how something \textit{should} work. 

Given this diversity in understanding and the possibility of misinterpretations, the need for a level of rigorousness is apparent. Ambiguity can be eliminated by formalizing a framework for defining and composing concepts. Of course, even a conceptual model could prove to be insufficient or unreasonable in practice, so integrating a mechanism for translating the model into a formal modeling language would be valuable. Specifically, a translation to a language with model finder support would allow developers to quickly analyze and explore their designs in great detail, possibly validating certain properties of the design. 

\newpage
\section{Goals}\label{sec:goals}
Building on the motivation that was outlined above, this thesis has aimed to reach the following goals:

\begin{enumerate}[label=\textbf{Goal \arabic*:},leftmargin=*,align=left]
    \item Develop a well-defined \gls{dsl} for concept-based design. In particular, the \gls{dsl} must define a way to represent concepts and should include language constructs for composing concepts. 
    \item The \gls{dsl} should be compilable to the Alloy modeling language  \cite{Alloy6Documentation,DanielJacksonAlloy} for automated and exhaustive analysis of a specification.  
    \item The syntax and semantics of the \gls{dsl} should be illustrated with several examples and reflect principles of concept-based design.  
    \item Custom language support could be added to Visual Studio Code. 
\end{enumerate}

Above the verbs \textit{must}, \textit{should} and \textit{could} denote different levels of importance following conventional \glsxtrshort{moscow} \cite{moscow_ITIL_service_design} prioritization. 

\section{Reading Guide}\label{sec:reading_guide}
Aiming to enhance understanding of the thesis, this section outlines conventions adopted throughout the thesis for presenting information.

Concrete code snippets of reasonable length, or other structures pertaining to code, that appear in this thesis are clearly designated in listings (see listing~\ref{lst:example}). 

\begin{lstlisting}[caption={[Example Listing] Example listing of a C-program.}, label={lst:example}, float=ht]
#include <stdio.h>
int main() {
   printf("Hello World!");
   return 0;
}
\end{lstlisting}

To enhance readability, language keywords in \textit{Conceptual} and Alloy will be emboldened without implying any difference in meaning. Moreover, when the thesis refers to concrete names from listings, \textit{Conceptual}'s codebase, particular file extensions, or keywords of other languages, this will be highlighted as \kw{main}, keeping the same capitalization. Notably, by convention, lexical tokens are written in all caps to make them easier to distinguish, e.g. \kw{COMMA}. Type signatures of functions will be presented similarly (see listing~\ref{lst:example_sig}). 

\begin{lstlisting}[caption={[Example Listing of Signature] Example listing of a function signature}, label={lst:example_sig}, float=ht]
a ->  b * c -> D.d1
\end{lstlisting}

In listing~\ref{lst:example_sig}, the arrow delimiter is used to distinguish the types of the different parameters, and the star is used to signify a tuple. Modules (think files) begin with a leading capital letter, and the period operator is used to access something within a module. As an example, if the function is not partially applied, listing~\ref{lst:example_sig} will return a value of type \kw{d1} defined in module \kw{D}. If any, higher-order functions will be explicitly marked 
by wrapping nested functions with parentheses. 

After \hyperref[toc]{the table of contents}, a \hyperref[toc:listings]{list of all listings} and a \hyperref[toc:figures]{list of all figures} are included. Conventions regarding how parsing rules are presented will not be expanded upon here for two main reasons. Firstly, even if a convention were enforced, it would predominantly be used in a single section pertaining to parsing - as opposed to throughout the thesis. Secondly, different notations are introduced. Instead, the necessary notation will be presented prior to its initial use in the relevant sections. 

Additionally, the thesis uses several acronyms. The first time acronyms are introduced, they generally include both the full meaning and the shortened form in parentheses, e.g., "\gls{api}." Afterward, the \gls{api}-shorthand will be used. A comprehensive list of all acronyms used can be found in appendix~\ref{appendix:acronyms}. In a few instances, acronyms will appear in their shorthand form from the outset, without an introduction. This applies in some cases where acronyms are introduced in the footnotes, have a vastly more common shorthand, or represent terms whose full forms are not widely known or used. The decision is made on a case-by-case basis to ensure maximum clarity. In any case, acronyms will always be included in appendix~\ref{appendix:acronyms}. Acronyms and citations can be clicked, taking the reader to the corresponding entry in the acronym list or the bibliography. Similarly, references to figures, code listings, and sections are linked to their respective place in the document.  

\newpage
\section{Structure}\label{sec:structure}
As suggested above, the thesis largely revolves around formalizing a \gls{dsl} for concepts and source-to-source translating this language to Alloy. An account of concepts and the Alloy modeling language is given in chapter~\ref{chap:background}. This chapter additionally describes basic compiler theory for the concrete phases utilized by the compiler. 

Subsequently, chapter~\ref{chap:compiler} details the architecture of the proposed compiler and how the choice of implementation language might affect its design. Moreover, the chapter presents the specific design methodology used to develop the compiler.

Then chapters~\ref{chap:lexing_conceptual} through \ref{chap:cg_code_generation} elaborate on the compiler phases outlined in the preceding chapter, introducing the employed tools and discussing technical details of lexing, parsing, semantic analysis, and code generation. 

Chapter~\ref{chap:cases} uses the \gls{dsl} to express several concepts and applications. For a few select instances, the compiler translates the models to Alloy, allowing them to be reasoned about using Alloy's analysis tools.

The last chapter, chapter~\ref{chap:concluding_remarks}, concludes the thesis by reflecting on the underlying work. It comments on whether the goals of this chapter have been accomplished. Moreover, it discusses the learning outcome, other concept-related work, and project shortcomings. 

Appendix~\ref{appendix:acronyms} contains a comprehensive list of all the technical acronyms and abbreviations used throughout the thesis. In appendix~\ref{appendix:lexical_tokens}, an overview is given of the concrete character sequences of each lexical token used by the parser. Appendix~\ref {appendix:complete_syntax} presents the complete language syntax based on the tokens outlined in appendix~\ref{appendix:lexical_tokens}. Finally, appendix~\ref{appendix:language_support_vscode} discusses the extent to which the \gls{dsl} is supported in Visual Studio Code. 

The entire codebase for this project is publicly available on GitHub \cite{conceptual-repo}.

\begin{figure}
    \centering
    \includegraphics[scale=0.20]{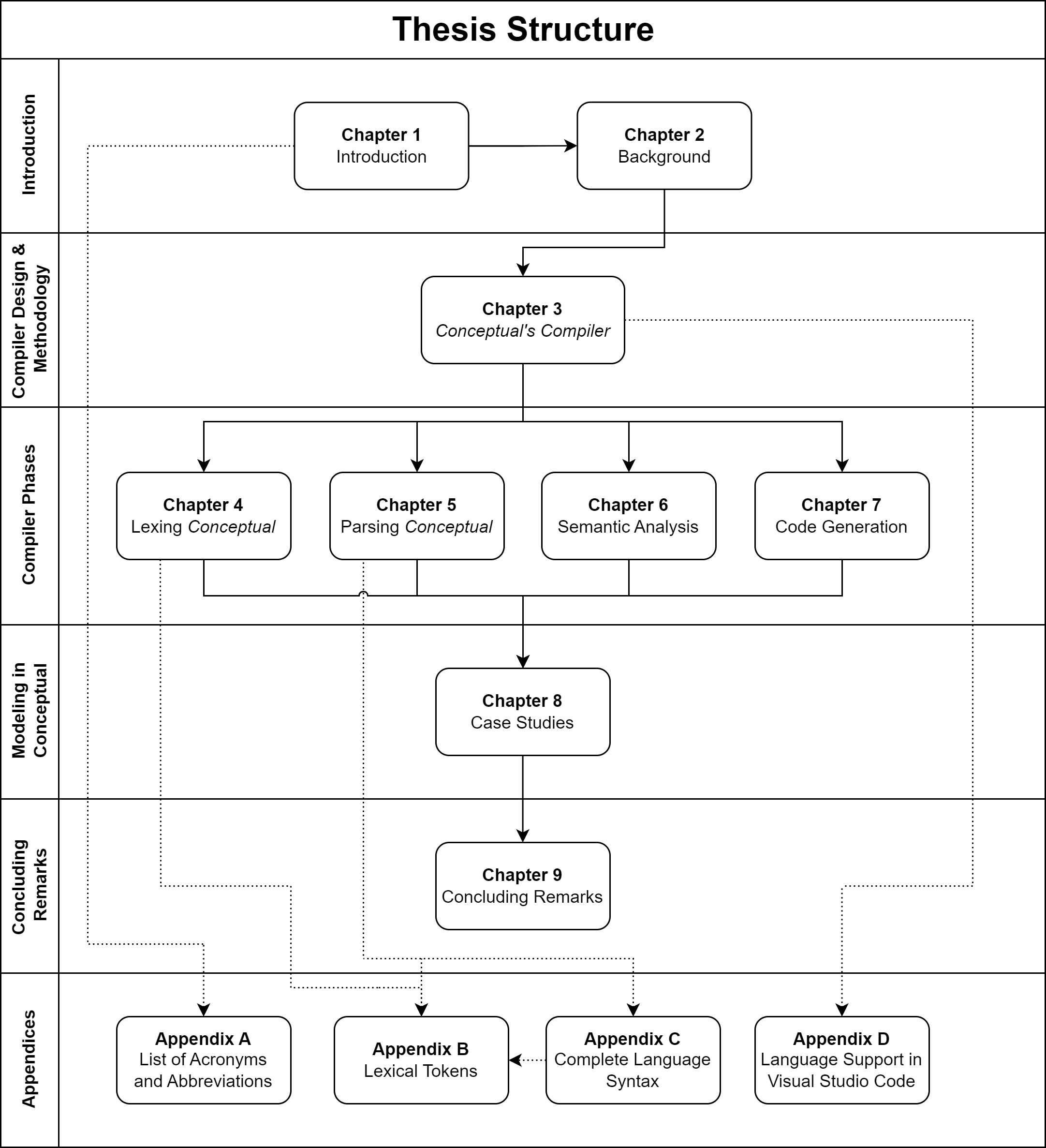}
    \caption[Graphical Representation of Thesis Structure]{Graphical representation of the thesis structure. Solid lines depict dependencies between chapters, whereas the dotted lines associate appendices with specific chapters.}
    \label{fig:thesis-structure}
\end{figure}

\chapter{Background}\label{chap:background}
\textit{This chapter briefly describes the core theory and the tooling used in the project. It describes the choice of programming language and tools, expands upon concepts, and presents the target language, Alloy, alongside relevant compiler theory. Specifically, section~\ref{sec:background_concepts} elaborates on the notion of concepts, and section~\ref{sec:background_related_work} presents existing concept-related work. Then section~\ref{sec:background_alloy} describes the fundamentals of Alloy. Finally, section~\ref{sec:compiler_theory} and its subsections (\ref{subsec:lexing_theory} through \ref{subsec:cg_theory}) briefly present the purposes of specific compiler phases and highlight the adoption of tools for certain phases.}

\section{Concepts}\label{sec:background_concepts}
As Daniel Jackson \cite{JacksonConcepts} puts it: 
\begin{quote}
    A concept is a particular solution to a particular design problem - not a large and vague problem but a small and well-defined need that arises repeatedly in many contexts. 
\end{quote}
That is, concepts are \textit{invented} by someone at some point in time to fulfill a particular purpose. Concepts should be one-to-one with this purpose: for every concept, there should only be one purpose that motivates it and vice-versa. In one sense, concepts are mental constructs that must be understood to use an application effectively. In a more formal sense, a concept can be considered a deterministic and coherent unit of functionality: given a state and a set of argument values for an action, at most one state can result from executing it, and consequently, an action cannot be refused arbitrarily \cite{JacksonConcepts}. Effectively, concepts can be modeled using a state machine with "\gls{csp}"-like synchronization \cite{hoare1985communicating}. And just like mental constructs have to be understandable without reference to other mental constructs, concepts are freestanding, defined by themselves without references to any other concept. Consequently, concepts offer a more stringent separation of concerns than traditional specifications \cite{JacksonConcepts}.  

In \cite{JacksonConcepts}, a concept definition includes its name, purpose, state, actions, and operational principle. A concept's operational principles can be viewed as defining archetypical scenarios that teleologically explain how the concept works and fulfills its intended purpose. These scenarios are very much akin to theorems about the behavior of the concept - that after some sequence of actions, a particular condition holds. They characterize the concept's essential behavior without detailing every aspect of its functionality. If there is no behavior or compelling purpose, there is no concept. The state can be summarized as the concept's internal memory. Actions describe the dynamic behavior of the concept. The combination of model state and atomic actions defines the behavior of the state machine. 

A system can be composed of several concepts. When a system integrates various concepts, it coordinates their actions without introducing new actions to any particular concept. Instead, it restricts the possible traces of actions that can occur. In such a system, each concept effectively runs autonomously, determining when its actions may occur and the impact of said actions on its internal state. A concept cannot directly alter another concept's state or influence the behavior of its actions, but its state can be observed \cite{JacksonConcepts}. Synchronizations can be used to constrain\footnote{Limiting the possible traces of actions in the system.} actions by forcing actions of one concept to trigger with certain aspects of another concept. Synchronizations are effectively the only way concepts can communicate, and since arguments to actions must be immutable, they help eliminate unintended interactions and side effects. The individual concepts of a composition can run in parallel \cite{JacksonConcepts}. 

\section{Other Work with Concepts }\label{sec:background_related_work}
The paradigm of concept-based design is still in its infancy, meaning there is only limited literature and work directly relevant to this thesis. Nevertheless, some preliminary research has been conducted with concepts in industry environments \cite{wilczynski2023concept_experience_report}. Additionally, a few notable projects utilizing concepts are described below. 

\textbf{Déjà Vu} \cite{DejaVu_thesis,DejaVu_tool} is an open-source platform for end-user development of apps with minimal code. It features a rich catalog of concepts, each representing a piece of full-stack functionality. After including and configuring concepts in \glsxtrshort{json}, they can be glued together using declarative bindings expressed in an \glsxtrshort{html}-like template language. Finally, a Déjà Vu app can be compiled into a standard \gls{mean} app. In collaboration with Daniel Jackson, Santiago Perez De Rosso, who spearheaded the development of Déjà Vu, did a complete redesign of \glsxtrshort{git} using conceptual design theory \cite{GitConceptualAnalysis}. The goal was to make the version control system less complicated for non-git-aficionados. This effort culminated with \textbf{Gitless} \cite{GitlessPaper}, which attempts to remedy well-known difficulties of \glsxtrshort{git}. 

Current research explores concepts in a less formal and rigorous setting than this thesis presents. Specifically, the \textit{concept-ai} team led by Daniel Jackson within the Software Design Group at \glsxtrshort{csail} is researching the application of concepts in \glspl{llm}. With Abutalib "Barish" Namazov as its lead architect and implementor, they are developing a tool \textbf{Kodless} \cite{kodless_repo}, which, by providing an \gls{llm} with concept specifications, can generate and automatically deploy full-stack web applications with \gls{crud} functions on a shared database. The prompts provided are used to generate an \gls{api} backend and a textual representation of a reactive frontend written in a custom \glsxtrshort{html}-inspired \gls{dsl}, which can then be interpreted. Using \glsxtrshort{gpt}-4 \cite{OpenAI_GPT4_2023}, they have had success generating clones of sites, such as Hacker News, by outlining the concepts, app compositions, and synchronizations in a few lines of natural language. 

% 1) Provide prompts that outline each concept
% 2) GPT expands upon this prompts, generating "concept code"
% 3) Briefly describe certain routes of the API
% GPT synthesizes the prompts in 3) into cdoe that calls the concept actions using the specs that was generated in 2).
% General insight: Limited context window, limited control if code generating all at once. Introducing a lot of layers appears to provide more control. 

\section{Alloy}\label{sec:background_alloy}
The Alloy modeling language \cite{Alloy6Documentation} is a formal and declarative specification language initially developed by Daniel Jackson and his group at \gls{mit}. Alloy has relational logic and set theory at its foundation, incorporating the quantifiers of first-order logic with the operators of relational calculus \cite{DanielJacksonAlloy,DanielJacksonSoftwareAbstractions}. By only allowing first-order structures, Alloy's operators can be defined in a very general way, which allows the language to remain lightweight \cite{DanielJacksonAlloy,JacksonAlloyLogicalModellingLanguage}. However, even first-order logic is undecidable \cite{turing1936aUNDECIDABLE}. The implication is that Alloy has no effective algorithm to correctly classify all statements as true or false. Considering the impracticality of a complete analysis, Alloy instead only considers a finite space with bounded model checking \cite{Biere1992BoundedModelChecking,Cunha_2014AlloyBoundedModel}, permitting infinite models. The usefulness of this approach relies on the fact that most bugs can be demonstrated with small counterexamples \cite{SmallScopeHypothesis}. Even in a small scope, however, the space is too large to search through explicitly \cite{DanielJacksonSoftwareAbstractions}. Alloy's front-end analysis tool, the Alloy Analyzer, instead compiles the problem into a \gls{sat} problem while also reducing the problem by applying a variety of strategies, e.g., adding symmetry-breaking constraints to remove equivalent cases \cite{DanielJacksonSoftwareAbstractions}. This effectively delegates the task of finding a witness\footnote{or showing that no such witness can exist} to off-the-shelf \gls{sat}-solvers such as Kodkod \cite{KodKod_torlak} and lets Alloy exploit recent advances in constraint-solving technology. While the analysis is incomplete, the search space is large enough to offer a degree of coverage that is unattainable by testing \cite{DanielJacksonSoftwareAbstractions}. Some of the key features of the analyzer are its ability to provide immediate visual feedback, check user-defined assertions, and generate instances of models that satisfy the specification when no counterexample is found.  

The Alloy language consists of several constructs generally recognizable to developers, including modules, polymorphism, and parameterized functions. However, it also introduces some unique features and keywords that distinguish it from other languages \cite{LausdahlVDM2Alloy}: 
\begin{itemize}
    \item A \textit{signature} \kw{sig} represents a set of atoms. Atoms are primitive entities with three fundamental properties: they are indivisible, immutable, and uninterpreted. A signature declaration can optionally be accompanied by a set of \textit{fields} representing a relation between atoms. 
    \item Model constraints that are assumed to always hold are defined by \kw{fact} clauses.
    \item An assertion using \kw{assert} expresses properties that are intended to follow from the facts of the model.
    \item A predicate \kw{pred} defines a formula and is similar to a traditional function that returns a boolean. Functions \kw{fun} share the same structure as predicates but return a value, effectively defining a way of getting an atom, a set, or a relation. Both predicates and functions can take arguments and generally cannot be recursive. 
    \item \textit{Commands} are instructions to perform a particular analysis. These commands include \kw{run} to search for an instance satisfying a given formula and \kw{check} that attempts to contradict a formula by searching for a counterexample. 
\end{itemize}

The latest version of Alloy (Alloy6) \cite{Alloy6Documentation} introduced an implicit, built-in notion of discrete time in the form of states. Alloy models using this internal clock are effectively represented as an infinite sequence of instances, also called a trace. In a trace, each instance corresponds to the state of a dynamic system where relations can change with the state transitions. With these changes came a set of 
\gls{ltl} operators \cite{LamportTemporalLogicOfActions}, allowing developers to express properties over time as properties over traces. These operators include:  \kw{after}, \kw{always}, \kw{eventually}, \kw{until}, \kw{historically}, \kw{releases}, and many more \cite{Alloy6Documentation}.

\section{Compiler Theory}\label{sec:compiler_theory}
The subsections below briefly describe the purposes of each compiler phase used in the final design (see figure~\ref{fig:compiler_architecture}). Additionally, some general insights are shared that advocate for using specific tools when applicable. 

\subsection{Lexing}\label{subsec:lexing_theory}
The goal of the lexer\footnote{Sometimes referred to as a scanner or tokenizer.} is to take a stream of characters, e.g. an input file, and transform it into meaningful sequences of \glsxtrshort{ascii} characters known as \textit{tokens} \cite{hoe1986compilers_dragon}. Lexers are generally written by hand or generated using existing tools \cite{FowlerDSL}. While hand-written lexers offer more control and can be optimized for specific uses, it is non-trivial and takes significant resources to implement a lexer by hand, even for a simple language \cite{FowlerDSL}. Lexer generators are normally preferred due to their ease of use, general efficiency, and flexibility. In fact, \cite{gray1988generator_lexer} shows that generators can produce executable code that can run as fast as hand-coded lexers. As such, \textit{Conceptual} uses a lexer generator.

\subsection{Parsing}\label{subsec:parsing_theory}
The goal of parsing is syntactical analysis, i.e.\ to systematically decompose a given sequence of tokens into its constituent structures, ultimately constructing an \gls{ast} \cite{hoe1986compilers_dragon}. This process is guided by a set of production rules in a \textit{grammar}, which define how tokens can be combined to form valid structures according to the language's syntax. The type of grammar needed depends on the underlying language and the parsing strategy \cite{Chomsky57_syntactic_structures,hoe1986compilers_dragon}. As was the case for the lexer, the choice is often between creating a parser by hand or using a parser generator. The main advantage of writing a recursive parser by hand is the ability to provide high-quality error messages on syntax errors and possibly also error recovery; the disadvantage is that it is more time-consuming and requires more code to implement than using generators \cite{FowlerDSL}. On the contrary, it is (usually) quite easy to quickly write a working and maintainable specification for a generator, but they rarely provide the same level of end-user friendliness. Syntactically incorrect inputs are frequently rejected with a generic parsing exception; some tools let you integrate specific error productions directly in the grammar, yet this approach can often lead to convoluted code \cite{MenhirReferenceManual2023}.

Today, generator tools are very mature and can handle complex languages efficiently \cite{FowlerDSL}. Regarding parsing a \gls{dsl}, \cite{FowlerDSL} highlights only the following drawbacks of using a generator: they have a steep learning curve, they complicate the build process, and good tools might not exist for the language platform. Consequently, \textit{Conceptual} uses a parser generator. 

\subsection{Semantic Analysis}
After verifying the syntactic correctness of the input and producing a complete parse tree, the next step is to ensure it is also logically coherent and semantically well-formed \cite{appel2004_modern_compiler_ml}. During this phase, the \gls{ast} produced by the parser is traversed, systematically checking each node for errors and annotating expressions with type information to produce a typed \gls{ast}. While explicit type annotations may not be strictly necessary for code generation to Alloy\footnote{But mandatory for other typed languages or \glspl{ir} such as \glsxtrshort{llvm}.}, the information they provide can still help identify type mismatches and other semantic inconsistencies more easily. 

\subsection{Code Generation}\label{subsec:cg_theory}
Once the input has been semantically verified, it can be translated into the target language. Crucially, the translation must preserve semantics, and the output must be semantically equivalent to the input \cite{hoe1986compilers_dragon}. This phase involves translating nodes from the typed \gls{ast} into corresponding constructs within the target language and serializing them.

\chapter{Conceptual's Compiler}\label{chap:compiler}
\textit{This chapter outlines the general structure of the proposed compiler and the methodologies behind its design. Concretely, section~\ref{sec:compiler_architecture} presents the compiler architecture. Then, section~\ref{sec:rationale_language} discusses the choice of implementation language and its potential impact on the final architecture. Afterward, section~\ref{compiler:sec_language_support} briefly comments on the tool support provided in Visual Studio Code. Finally, section~\ref{sec:compiler_method} presents the methodology that was used for designing the compiler. For reference, the concrete compiler phases are discussed in detail in chapters~\ref{chap:lexing_conceptual} through \ref{chap:cg_code_generation}.
}

\section{Compiler Design and Architecture}\label{sec:compiler_architecture}

The literature tells us that the field of language and compiler design is vast; entire books are dedicated to individual compiler passes or aspects of them \cite{AdvancedCompiler_Muchnick}. Nevertheless, compiler connoisseurs and language lovers can be very \textit{opinionated} due to the numerous decisions and trade-offs that can significantly impact efficiency, functionality, and usability. Consequently, many approaches have different costs and benefits \cite{hoe1986compilers_dragon,scott2000language_compiler,AdvancedCompiler_Muchnick}. When designing a \gls{dsl}, the decision to develop an external or an internal (embedded) \gls{dsl} significantly influences the approach one should take \cite{FowlerDSL}. In this project, an external \gls{dsl} is built. This choice is largely motivated by the proposal of an external syntax in \cite{JacksonConcepts}. However, this choice has other merits beyond this, such as greater domain-specificity, isolation from a host language, and complete autonomy over the tooling environment, syntax, and semantics. A consequence of this choice is that a custom lexer and parser are generally required to interpret the language \cite{FowlerDSL}. If the language is simple, delimiter-directed translation or standard string processing techniques may be sufficient \cite{FowlerDSL}. However, syntax-directed translation using a formal grammar is often preferable. For this project in particular, the adopted syntax supports hierarchical structures, is whitespace insensitive, and generally omits delimiters. These characteristics make approaches like delimiter-directed translation impractical, if not impossible. As such, this project implements the lexer and parser as separate entities, providing a more stringent separation of concerns. Semantic analysis is then used to ensure the language constructs are used correctly and meaningfully. The rationale behind including an explicit phase for code generation was touched on in section~\ref{sec:rationale_language}.

\begin{figure}[H]
    \begin{adjustbox}{center}
    \input{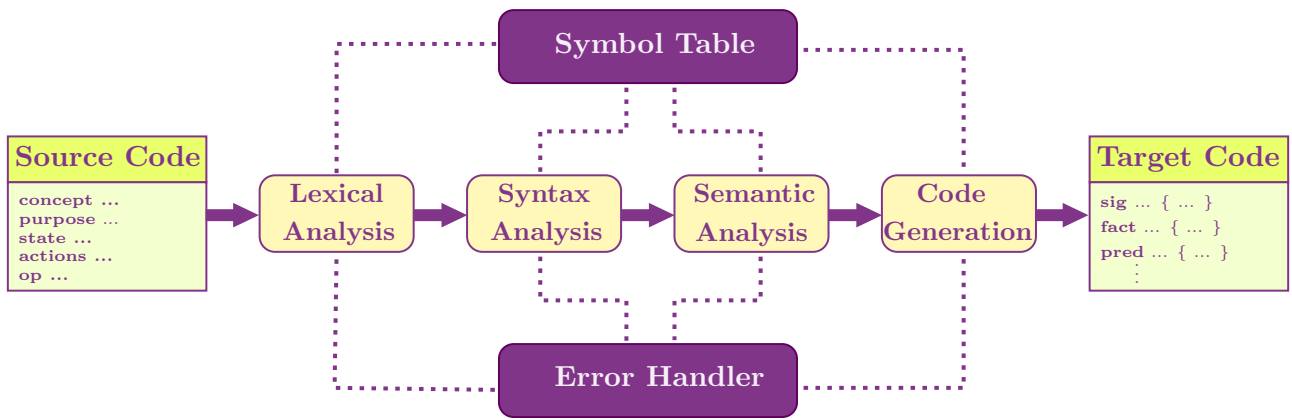}
    \end{adjustbox}
    \caption[Compiler Architecture]{Overall compiler architecture and phases.}
    \label{fig:compiler_architecture}
\end{figure}

The overall architecture of the compiler is depicted in figure~\ref{fig:compiler_architecture}, and each pass will be individually expanded upon in the subsequent sections. The overall structure resembles that of a compiler for general-purpose languages as seen in textbooks \cite{appel2004_modern_compiler_ml,hoe1986compilers_dragon}, only missing the translation to an \gls{ir} and code optimization. In the proposed pipeline, emphasizing these phases makes little sense as the target is a modeling language - not the final implementation. Hence, performance considerations are less relevant, mostly affecting the scope size that can feasibly be analyzed. 

\section{Choice of Implementation Language}\label{sec:rationale_language}
There have always been proponents for a particular programming style \cite{gries_programming_methodology}. Features from functional languages like pattern matching and immutability are incredibly useful for dealing with variant types and the notion of scope \cite{appel2004_modern_compiler_ml}, making such languages popular for compiler design. Still, many designers prefer an imperative and object-oriented style, which can generally achieve the same thing \cite{FowlerDSL} with patterns like the visitor pattern. 

In terms of concrete languages, the obvious imperative choice is Java. Java has a massive ecosystem and has even gradually integrated a lot of the features native to the functional paradigm: immutability with \kw{final}, higher-order functions through lambda expressions in Java 8, and it is possible to emulate a form of pattern matching with the \kw{instanceof} operator introduced as a preview feature in Java 14. The principal advantage, though, would be the fact that the Alloy language is written entirely in Java \cite{Alloy6Documentation,AlloyRepo}. It would then, on paper, be possible to avoid proper code generation by creating objects conforming to the Alloy \gls{ast} and calling the Alloy \gls{api} directly. However, this would not be without costs. Specifically, the complexity would move from code generation to the parsing and semantical analysis phases, shifting the focus from the Alloy language to its internal representation. Alternatively, languages like Kotlin or Scala, which runs on the \gls{jvm} and compiles to \gls{jvm} bytecode, could also avoid explicit code generation while having access to the entire \gls{jvm} ecosystem. Additionally, both languages embrace features such as first-class functions from the functional paradigm. Nevertheless, both languages are relatively verbose, and Scala, in particular, has lengthy compile times \cite{krill_jetbrains_2011_scala_jetbrains}.

Although \gls{jvm} languages are also a viable option for source-to-source translation, \textit{Conceptual} uses OCaml. OCaml is a dialect of the \gls{ml} and supports all the features characteristic of a functional language, including first-class functions and pattern matching. Additionally, its strong and static type system\footnote{Based on Hindley-Milner type inference from lambda calculus.} combined with a lack of overloading result in sophisticated type inference that outshines the capabilities of Java 10's generics and \kw{var}. The type system natively supports parametric polymorphism while being written in an entirely untyped style\footnote{Although some people prefer writing explicit type annotations for top-level functions.}. Consequently, it is possible to write concise and readable code, particularly for variant types with pattern matching and recursion\footnote{Tail-recursion is also optimized by the OCaml compiler.}. This reduces complexity and increases maintainability and extensibility. Many functional languages share these characteristics, and so the choice of OCaml is influenced by prior experience in addition to its history of designing industry-grade languages\footnote{For instance, the Coq proof assistant, the Hack programming language by Facebook to extend its \glsxtrshort{php} codebase, or the static type checker Flow for \gls{js} also by Facebook employees.}. Specifically, the most recent major release of OCaml is used, version 5.0.0, which introduced a new runtime system, allocator, and garbage collector. 

The choice of implementation language has no impact on the architecture of the compiler. This remains true except possibly if \gls{jvm}-languages are used, in which case one \textit{can} bypass an explicit code generation phase. However, as the syntax of the \gls{dsl} (see chapter~\ref{chap:parsing_conceptual}) largely coincides with the syntax of Alloy, proper source-to-source translation is not overly complicated. The main effect of the language is then on the ecosystem and the available tools. Even so, many tools, especially for lexing and parsing, share a similar syntax regardless of the host language.

This project uses the Dune build system \cite{duneBuildSystem} (version 3.14), which is commonly used in OCaml development. Notably, OCaml has two compilers: a native compiler and a bytecode compiler \cite{ocaml_github}. The native compiler generally performs more optimizations, producing faster executables, and thus Dune uses this compiler by default when supported by the environment. The native compiler is also used in this project. 

\section{Language Support in Visual Studio Code}\label{compiler:sec_language_support}
To enhance the experience of modeling with \textit{Conceptual} and using the associated compiler, some work has gone into establishing language support in the Visual Studio Code \gls{ide}. However, due to the \textit{could} prioritization of this task, it will not be discussed in detail, and many quality-of-life features, which developers have come to expect, are omitted because they rely on external language servers. The level of support provided is elaborated on in appendix~\ref{appendix:language_support_vscode}.

\section{Methodology for Design of the Compiler}\label{sec:compiler_method}
Software quality tends to be positively affected when working in a structured manner \cite{Ashrafi2003impact_process_quality}. The process of designing a compiler for a custom language is intricate and, in turn, also generally benefits from working in a systematic way. As computer-based languages often evolve at rapid rates to keep up with user needs \cite{urma2017PL_evolution}, especially early in their lifetime, this project adopts an iterative approach for meeting changing demands and requirements \cite{kumar2012impact_agile}. Especially in compiler design, where insufficiencies may necessitate simultaneous adjustments in various phases, the ability to go back and immediately modify or refine preceding work is crucial. The adopted method ensures that each development phase is not only well-defined but subject to continuous testing and evaluation at various stages, improving robustness and error locality. The concrete methodology is highlighted on the next page. 

\newpage
\subsection{Concrete Methodology}
\begin{enumerate}
    \item Devise a strategy for input tokenization.
    \item Define rules for combining tokens into a coherent language syntax.
    \item Evaluate whether the defined syntax is sufficiently rich based on the criteria below.
    \begin{enumerate}
        \item There are no shift-reduce or reduce-reduce conflicts. If there are ambiguities, they must be dealt with by revising the grammar or by explicitly declaring precedence or associativity of certain tokens. 
        \item The syntax is general and flexible enough to accommodate a wide variety of scenarios. If the syntax is inadequate, modify the grammar and possibly the underlying tokenization scheme. 
    \end{enumerate}
    \item Define the semantics of the \gls{dsl} and what constitutes erroneous programs. 
    \item Evaluate the semantics according to the criteria listed below. If inadequacies are found, revise the semantics and, if necessary, revisit the previous steps. 
    \begin{enumerate}
        \item The established static semantics are sufficient for catching errors and helping ensure program correctness. 
        \item The mechanism for identifying semantic errors, such as type mismatches, is effectively designed and conveys information in a user-friendly manner. 
    \end{enumerate}
    \item Determine a plausible translation scheme to Alloy. 
    \item Evaluate the proposed mapping strategy to Alloy, focusing on the key aspects listed below. If the translation is satisfactory, the compiler is considered functional. If not, revise the translation and possibly revisit previous steps. 
    \begin{enumerate}
        \item Whether \gls{dsl} constructs are faithfully represented in Alloy and that the translation preserves semantic integrity.
        \item The practicality and ease of use of conducting further analysis using Alloy's model finder.
    \end{enumerate}
\end{enumerate}

\newpage
\subsection{Graphical Representation of the Methodology}\label{subsec:compiler_graphical_methodology}
\begin{figure}[ht!]
    \centering
    \includegraphics[scale=0.25]{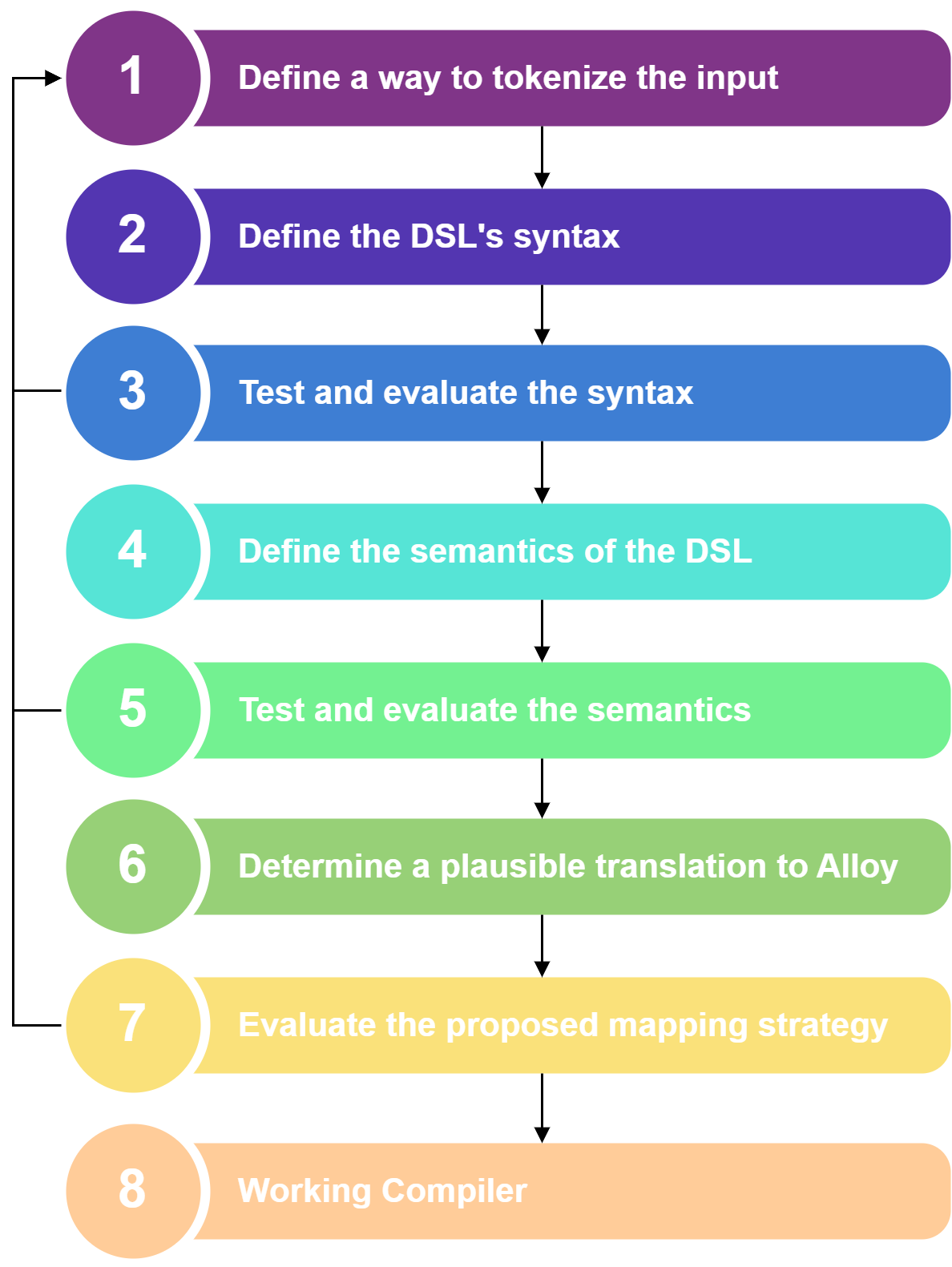}
    \caption[Graphical Representation of Design Method]{Graphical representation of the compiler design methodology from section~\ref{sec:compiler_method}.}
    \label{fig:comp_method_graph}
\end{figure}

Figure~\ref{fig:comp_method_graph} provides a graphical representation that encapsulates the overall design process based on the methodology steps above. This depiction simplifies the methodology, omitting most of the accompanying text. Furthermore, the graph is shown to always loop back to the first initial step. This cyclical nature does not imply starting from scratch but rather that \textit{any} preceding phase may be refined as part of the incremental design or due to an evolving understanding of the compiler's needs. Chapters dedicated to certain aspects of the proposed compiler will include parts of the figure in their prelude, indicating which step(s) the chapter discusses. 

\chapter{Lexing Conceptual}\label{chap:lexing_conceptual}
\textit{In this chapter, Conceptual's lexer is introduced. Structurally, section~\ref{sec:lexing_ocamllex} first introduces the concrete lexer generator used and the rationale behind its configuration for Conceptual. Then section~\ref{sec:general_lexing_strategy} introduces the overarching lexing strategy. Finally, section~\ref{sec:lexing_reference_conceptual} details the lexer's concrete behavior at a reference level. }

\includegraphics[width=\textwidth]{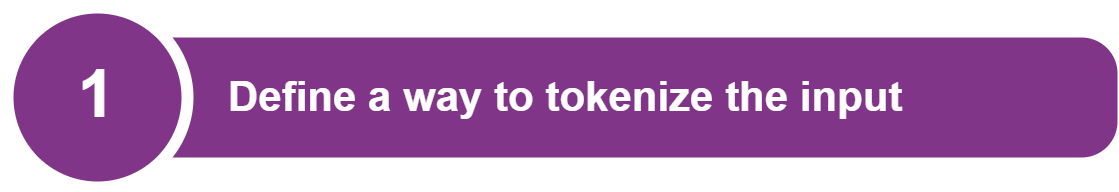}

\section{Ocamllex}\label{sec:lexing_ocamllex}
As its lexer generator \textit{Conceptual} uses \textit{ocamllex} \cite{OCamlManual2023LexerParser}, which is inspired by \textit{Lex} \cite{Lesk90_Lex} and essentially operates as a regex-based table-driven lexer that supports multiple mutually recursive lexical states. Provided an \kw{.mll} specification file (see listing~\ref{lst:lexing_structure}), which fundamentally is similar to a set of regular expressions of the form $R_1 | R_2 | ... | R_n$ where $R_i$ is the description for token $i$, the tool constructs a \gls{nfa} \cite{Rabin59_NFA} from the regular expressions.

\begin{lstlisting}[caption={[Lexer Specification Structure] General structure of the lexer specifications. Here \kw{parse} denotes that the "longest match" rule applies: it selects either the longest prefix of the input or, in the case of ties, the rule that appears earlier. The \kw{shortest} keyword does the opposite. Headers and trailers can include ocaml code that will be copied verbatim into the generated artifact.}, label={lst:lexing_structure}, float=ht, keywordstyle=,]
{ header }
let identifier = regex ...
rule entrypoint [arg1... argn] = parse 
    | regex { action }
    | ...
    | regex { action }
and entrypoint2 [arg1... argn] = shortest
    | regex { action }
    | ...
and ...
{ trailer }
\end{lstlisting}
In practice, however, a \gls{dfa} \cite{kleene1956representation_FA_RE,mcculloch43pitts_nerve_nets_FA_prior} is more efficient for the actual matching process\footnote{Matching with \glspl{nfa} involves backtracking, thus requiring more work} and so the \gls{nfa} is compiled into a \gls{dfa} using algorithms like subset construction \cite{appel2004_modern_compiler_ml}. The resulting \gls{dfa} is then minimized as it can be exponentially large. Finally, OCaml code is generated that implements the \gls{dfa}, including a function \kw{lex} that reads the input text and executes the associated action for each recognized token. By default, the output code uses OCaml's built-in automata interpreter. For the bytecode compiler, this is the most optimal method. However, the native compiler used in this project is more efficient when adding the \kw{-ml} flag, which encodes the automaton as OCaml functions \cite{OCamlManual2023LexerParser}.

\section{General Lexing Strategy}\label{sec:general_lexing_strategy}
\textit{Conceptual} universally resolves lexical ambiguities by using the \textit{longest match} rule and opts to use stream-based lexing techniques. In the OCaml ecosystem, this means producing a lexer of the form seen in listing~\ref{lst:lexer_signature}. 

\begin{lstlisting}[float=ht, caption={[Signature of Lexer]Signature of stream-based lexer in ocamllex},label={lst:lexer_signature}]
    Lexing.lexbuf -> Parser.token
\end{lstlisting}

The function in listing~\ref{lst:lexer_signature} is then invoked procedurally by the parser to generate tokens as needed. Here \kw{Lexing} denotes the standard run-time module for lexers \cite{OCamlLexingModule2023} and \kw{lexbuf} is a data structure within this module that represents the current state of the lexical analyzer. As a general remark, the lexing engine modifies the \kw{lexbuf} in-place and assumes that users do not touch this structure. The \kw{Parser} module is the artifact produced by the parser generator and contains a variant type of all possible tokens. Having this dependency ensures that tokens produced by the lexer are consistent with the ones used by the parser. 

Batch tokenization is a viable alternative to stream-based approaches, especially in cases where additional analysis and manipulations on the tokens may be beneficial before parsing, but this is normally excessive for a \gls{dsl} \cite{FowlerDSL}. Moreover, it is still possible to slightly modify a stream-based lexer. In fact, \textit{Conceptual} extends the generated lexing artifact with a \kw{TokenCache} module as a kind of middleware through which additional tokens can be injected into the stream. The parser first retrieves tokens from the cache, resorting only to using the lexer to consume text input and emit new tokens when the cache is depleted. This feature was used sparingly to simplify the parsing phase (discussed briefly in chapter~\ref{chap:parsing_conceptual}). For conciseness, most dependencies (e.g. error, cache, and parsing modules) alongside helper logic for strings, integer overflows and name mangling are included in the specification prelude, which is copied verbatim into the generated lexer. 

In \textit{Conceptual}, most of the lexing rules are confined within a single lexical state called \kw{lex}. However, there are separate lexical states for handling string data types, comments, and nested comments. Upon recognizing an unspecified character or sequence of characters, the lexer immediately throws an exception that contains the violating string.

\section{Conceptual Lexing Reference}\label{sec:lexing_reference_conceptual}
This section describes concrete lexical aspects specific to \textit{Conceptual}, primarily concerned with the different categories of tokens: reserved keywords, identifiers, literals, operators, and punctuation symbols. The subsections below are meant to be brief and provide a reference-level overview of these areas. 

\subsection{Reserved Keywords}
The following character sequences are reserved and may not be used as identifiers. 

{\scriptsize
\begin{multicols}{6}
\noindent
\texttt{and} \\
\texttt{or} \\
\texttt{when} \\
\texttt{can} \\
\texttt{until} \\
\texttt{then} \\
\texttt{no} \\
\texttt{is} \\
\texttt{empty} \\
\texttt{none} \\
\texttt{in} \\
\texttt{not} \\
\texttt{set} \\
\texttt{one} \\
\texttt{lone} \\
\texttt{some} \\
\texttt{const} \\
\texttt{string} \\
\texttt{int} \\
\texttt{concept} \\
\texttt{purpose} \\
\texttt{state} \\
\texttt{actions} \\
\texttt{principle} \\
\texttt{app} \\
\texttt{include} \\
\texttt{sync} \\
\end{multicols}
}

\textit{Conceptual} is case-sensitive and distinguishes between upper- and lowercase letters. Notably, there are no keywords for "true" and "false"\footnote{The lack of boolean literals is expanded upon in section~\ref{sec:conceptual_parsing_ref}.}.

\subsection{Identifiers}
An \textit{identifier} is a sequence of letters, digits, and underscores beginning with a letter and is not a reserved word \cite{CLU-manual}. Identifiers are case-sensitive. Unlike some languages (Alloy included) \textit{Conceptual} does not support identifiers with a leading underscore, allowing leading underscores to be used for name mangling of identifiers representing reserved keywords in Alloy. 

\subsection{Literals}\label{subsec:lexing_literals}
The string and integer primitive types have literals. Integer literals are represented with 64-bit signed integers. String literals are encapsulated in quotation marks and have escaped character sequences translated into their meanings. The empty set literal is also supported using "\{\}", "none", or "empty". 

\subsection{Operators and Punctuation Symbols}\label{subsec:lexing_punctuation_operators}
\textit{Conceptual} reserves the following character sequences as symbols for operators and punctuation

\begin{center}
    
{\scriptsize
\begin{multicols}{8}
\noindent
\texttt{=} \\
\texttt{+} \\
\texttt{-} \\
\texttt{\&} \\
\texttt{:} \\
\texttt{;} \\
\texttt{.} \\
\texttt{,} \\
\texttt{!} \\
\texttt{\~} \\
\texttt{\^} \\
\texttt{\#} \\
\texttt{*} \\
\texttt{\%} \\
\texttt{\slash} \\
\texttt{(} \\
\texttt{)} \\
\texttt{[} \\
\texttt{]} \\
\texttt{\{} \\
\texttt{\}} \\
\texttt{|} \\
\texttt{"} \\
\texttt{<} \\
\texttt{>} \\
\texttt{<=} \\
\texttt{>=} \\
\texttt{\&\&} \\
\texttt{||} \\
\texttt{->} \\
\end{multicols}
}
\end{center}

\subsection{Comments and Other Separators}\label{subsec:lexing_comment_seperator}
Comments are sequences of characters ignored by the compiler. \textit{Conceptual} uses C-style syntax for commenting: characters between $//$ and a newline character, and from $/*$ to $\backslash*$, are treated as comments. Additionally, comments in the latter form can be nested. 

A \textit{separator} is either a comment or a blank character (space, horizontal tab, carriage return, and newline). Zero or more separators may generally appear between any two tokens.

\chapter{Parsing Conceptual}\label{chap:parsing_conceptual}
\textit{This chapter details various aspects of Conceptual's parser. First, section~\ref{sec:menhir} introduces the used parser generator and discusses its configuration options. Then the general representation of models is briefly presented in  section~\ref{sec:model_representation}, and section~\ref{sec:syntax_errors} outlines how syntax errors are handled. Afterward, in section~\ref{sec:example_grammar_rule}, one of Conceptual's grammar rules is analyzed as an example. Penultimately, similar to the lexing reference in section~\ref{sec:lexing_reference_conceptual}, section~\ref{sec:conceptual_parsing_ref} provides an overview of how different areas of the \gls{dsl} are parsed. Finally, section~\ref{parsing:sec_testing} briefly describes how the parser is tested.}

\includegraphics[width=\textwidth]{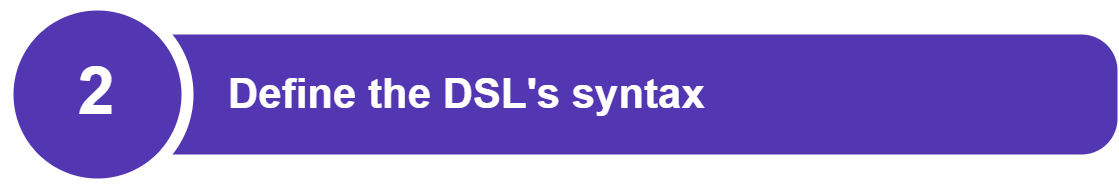}
\includegraphics[width=\textwidth]{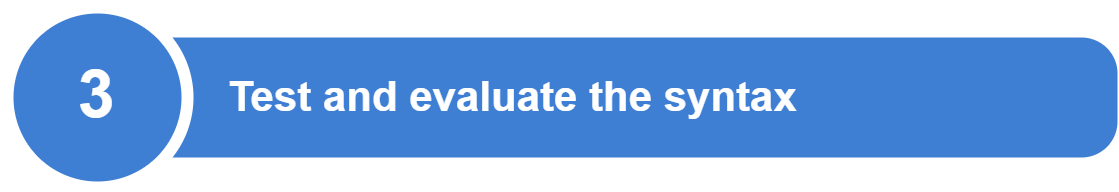}

\section{Menhir}\label{sec:menhir}
As its parser generator \textit{Conceptual} uses \textit{Menhir} (version 3.0) \cite{MenhirReferenceManual2023}, which integrates directly with ocamllex and has previously been used in the industry \cite{HackLang}. Concretely, Menhir is based on bottom-up LR(1)\footnote{The "LR(k)" notation means "translatable from left to right with a bounded lookahead of k"} parser construction techniques presented in \cite{KNUTH1965607_LR} but leverages default reductions and merges states to become practical. It is heavily inspired by its predecessors: \textit{\glsxtrshort{yacc}} \cite{Johnson1978YaccYA}, \textit{ML-yacc} \cite{MLYaccManual2000}, and \textit{ocamlyacc} \cite{OCamlManual2023LexerParser}. Menhir supports several types of back-ends: one is table-driven, one is code-oriented, and one uses Coq \cite{Huet2005TheCP_COQ} to produce proof that the parser is complete and correct according to the grammar. For the scope of this project, the latter is overkill. The \textit{table} back-end has a dependency to \kw{MenhirLib} and is significantly slower (2-5x) than the \textit{code} back-end \cite{MenhirReferenceManual2023}. Thus, the \textit{code} back-end was chosen for this project. Similarly to the lexer, this back-end compiles the LR pushdown automaton directly into a set of mutually recursive OCaml functions. The only deficit with this approach, as opposed to table-driven parsing, is that the code back-end only supports Menhir's monolithic \gls{api}. This \gls{api} embeds a \textit{single} top-level function, which is named after the entry point rule, into the parsing artifact (see listing~\ref{lst:menhir_signature}). \\
\begin{lstlisting}[caption={[Menhir's Monolithic Parsing Function] Signature of Menhir's monolithic parsing function in \textit{Conceptual}.},label={lst:menhir_signature}]
(Lexing.lexbuf -> Parser.token) -> Lexing.lexbuf -> Ast.model
\end{lstlisting}

This function takes two arguments
\begin{enumerate}[label=\arabic*)]
    \item a lexing function for producing tokens (commonly the lexing artifact)
    \item a \kw{lexbuf} representing the source from which lexing should start (such as the beginning of an input file) 
\end{enumerate}

and returns the root of the \gls{ast} representing the model, $\kw{Ast.model}$. The problem with the monolithic approach is that parsing specifications tend to only outline what constitutes a syntactically valid model. When given an erroneous model, the parsing artifact cannot infer the exact type of error that occurred, only that none of the parsing rules applies. Consequently, it immediately throws a generic exception when it cannot interpret the input. This contradicts the goals of a \gls{dsl}, particularly regarding end-user-friendliness. Menhir supports a special \kw{error} token, which is not produced by the lexer but can be incorporated into the grammar production rules. This allows better error reporting and possibly error recovery in certain cases. Unfortunately, this feature both pollutes the grammar and has been flagged for deprecation \cite{MenhirReferenceManual2023}. Alternatively, the table-driven backend supports an iterative \gls{api}, where the developer drives the parser and produces tokens. This requires additional source code, though not insurmountable, but lets the developer detect and handle individual errors as desired. The developer must predetermine the "error states" at the automaton level and then, for each, prepare a special diagnostic description in a special \kw{.messages} file format. This is non-trivial to do and requires a significant amount of bookkeeping. Especially since the number of erroneous states could foreseeably outnumber the number of correct ones. 

The generator must be provided with a context-free grammar specification to produce a parsing artifact. Specification files have the \kw{.mly} file extension, and they follow the syntactic structure seen in listing~\ref{lst:parsing_structure}. The header and trailer contain OCaml code copied verbatim into the parsing artifact. The declarations define the token type\footnote{This was also used directly by ocamllex to produce tokens.} including associativity and precedence rules for certain tokens. Optionally, the return type of each nonterminal symbol can be declared\footnote{This is mandatory for the code back-end. However, one can also invoke the bytecode compiler and rely on it to infer the types automatically. \textit{Conceptual} does the latter, and it has become the default behavior in newer versions of Dune.}. Finally, it is required to include a declaration of the parsing process's starting rule, the parse tree's root.

\begin{lstlisting}[caption={[Parser Specification Structure] General structure of the parser specification. }, label={lst:parsing_structure}, float=ht, keywordstyle=,]
%{
  header 
%}
  declarations
%%
  rules
%%
  trailer    
\end{lstlisting}

To define the syntax rules, Menhir supports an extended version \cite{MenhirReferenceManual2023,Scowen1998ExtendedBNF} of the standard \gls{bnf} grammar attributed to John Backus and Peter Naur \cite{Backus1959TheSA_BNF_v1,BackusNaur60_ALGOL}. The general form of a production is:

\begin{lstlisting}[caption={[Syntax for Parsing Rules] Syntax for the individual production rules in the grammar.}, label={lst:parsing_rule_syntax}, float=ht, keywordstyle=,]
nonterminal :
| symbol ... symbol { semantic-action }
| ...
| symbol ... symbol { semantic-action }
\end{lstlisting}

Tree construction occurs via the \textit{semantic actions}, which are fragments of OCaml code that execute when their associated production rule activates. The following extensions to traditional \gls{bnf} notation can be used in the specification

\begin{myalign}
$a?$ \qquad & for an optional a; \\
$a+$ \qquad & for a non-empty sequence of $a$; \\
$a*$ \qquad & for any sequence (including empty) of $a$.
\end{myalign}

Other syntactic shorthands exist, such as defining lists separated by one or more symbols. The parsing engine recognizes such operators and parameterized rules and expands them into an \textit{ordinary} \gls{bnf} grammar format before the parser is generated, making it possible to write concise grammars with fewer production rules.

\section{Model Representation}\label{sec:model_representation}
The goal of parsing, as mentioned in section~\ref{subsec:parsing_theory}, is to produce the \gls{ast} representation of a given model. 
The \textit{Conceptual} \gls{ast} defines the model type to be a record consisting of two lists: a list of \kw{concept} definitions and a list of concrete applications \kw{app} that compose the concepts. Either of these lists can be empty. 

\section{Syntax Errors in Conceptual}\label{sec:syntax_errors}
As alluded to in sections~\ref{subsec:parsing_theory} and \ref{sec:menhir}, Menhir's monolithic API is used, complicating the process of generating user-friendly error messages on syntax errors. 
Instead, upon encountering exceptions, \textit{Conceptual} uses the state of the lexer to approximate the location of the syntax error. This is mostly spot-on, but the accuracy depends on the granularity of the production rules - sometimes an error might happen on the subsequent line but never before the reported approximation. This suffices for two main reasons: 1) Concepts are defined to be small atomic units, and 2) \gls{dsl}s tend to have fewer language constructs than general-purpose languages, making errors easier to locate. Additionally, the speed-up achieved from using Menhir's code backend is non-negligible, seeing that the compiler omits an optimization phase, which is often the costliest component of a compiler. In a scarce number of cases, it is also still possible to produce well-formed syntax error messages.

\section{Production Rule Example}\label{sec:example_grammar_rule}
Using the Menhir notation introduced in section~\ref{sec:menhir}, listing~\ref{lst:EBNF} shows an example of a production rule from \textit{Conceptual} that uses $'+'$ to keep the logic for recognizing a concept's actions succinct. In this example, it recognizes the \kw{ACTIONS} token followed by at least one action\footnote{A concept without actions fulfill no purpose.} and then the token for operational principle \kw{OP}. When recognizing these, the semantic action dictates that it should construct an \kw{Actions} node, which is a record containing a list \kw{actions} of all the actions and a field with its location in the input file. The latter location attribute is included in essentially all nodes and used for error reporting. The syntax \kw{\$2} returns the contents stored in the second symbol, corresponding to the list of actions. Similarly, \kw{\$loc} is a special variable referring to the location of the input that is being parsed. 

\begin{lstlisting}[caption={[Example of EBNF Syntax] An example of a production rule from \textit{Conceptual}.}, label={lst:EBNF}, float=ht, keywordstyle=,]
c_actions :
| ACTIONS action+ OP { Actions{actions = $2; loc = mk_loc $loc} }
\end{lstlisting}

Moreover, the example highlights that a particular order is enforced for the concept constituents. Specifically, the elements of a concept must be defined in the order seen below:

\begin{verbatim}
    signature -> purpose -> state -> actions -> operational principle
\end{verbatim}

This ordering is logically coherent, as actions generally depend upon the state, and the operational principle may incorporate both the state and actions. Certainly, the parser could be extended to accommodate any permutation of these elements, but this would increase the complexity significantly and result in a much longer and less readable parser - all while being more counterintuitive. 

The delimiter-less and whitespace-insensitive nature of \textit{Conceptual} complicates the parsing process. Identifying where statements end and tokens belong becomes particularly challenging without conventional delimiters like semicolons, curly braces, or indentation to guide the structuring. The example above additionally showcases how the parsing process is simplified by using concept-related tokens\footnote{excess of which are generated by the token cache} as delimiters. 

\section{Conceptual Parsing Reference}\label{sec:conceptual_parsing_ref}
The following section describes various aspects of \textit{Conceptual}'s grammar entirely in terms of the underlying tokens for each rule. While the full set of productions will generally be omitted from this section, appendix~\ref{appendix:complete_syntax} contains the complete syntax for reference. However, a side effect of describing production rules in terms of their tokens is that \textit{Conceptual's} multi syntax does not show. For example, boolean operators allow the typical programming language style operators and the respective English words. The underlying strings for each token are defined in appendix~\ref{appendix:lexical_tokens}, which, by extension, identifies the tokens with more than one syntax. The grammar is explained using the standard \gls{bnf} operators:

\begin{myalign}
$a^*$ & for zero or more repetitions of $a$; \\
$a^+$ & for one or more repetitions of $a$; \\
$a\, |\, b$ & for $a$ choice of a or b; \\
$[a]$ & for an optional $a$.
\end{myalign}

In addition,

\begin{myalign}
$a_{sep}^*$ & means zero or more occurrences of $a$ separated by $sep$; \\
$a_{sep}^+$ & means one or more occurrences of $a$ separated by $sep$;
\\
$\epsilon$ & denotes the empty string.
\end{myalign}

Nonterminals are lowercase and encapsulated in $\langle\quad\rangle$. Lexer-produced tokens are uppercase and generally map directly to the character sequences outlined in appendix~\ref{appendix:lexical_tokens} but most represent idiomatic abbreviations, e.g. "concept" becomes \kw{CONCEPT} and ',' becomes \kw{COMMA}. The fact that tokens are used in the production rules, as opposed to character sequences, emphasizes that the parser is decoupled from the lexer; the underlying regular expressions for each token can be modified without changing the parser. The subsections below will discuss certain key language features using this notation. 

\subsection{Signature}
The entry point to defining a concept is the signature. The signature consists of the underlying name of the concept and can optionally be parameterized:

\begin{obeliskgrammar}
\gramfunc{\cUNDERSCOREsig{}}& \gramdef & \CONCEPT*{} \IDENT*{}                         \gramopt*{\LBRACK*{}                            \gramseplist{\COMMA*{}}{\IDENT*{}}               \RBRACK*{}}
\end{obeliskgrammar}

A parametric polymorphic signature is instantiated with its concrete generic types when synchronized with other concepts. It is syntactically valid to include brackets for a signature without parameters; however, the brackets are meaningless in this case. 

\subsection{State}
Most concepts require a notion of memory to be effective. The \textit{state} of a concept represents a set of mutable variables. The scope of these variables is from their declaration until the end of the enclosing concept. A variable declaration has the form:

\begin{obeliskgrammar}
\gramfunc{\cUNDERSCOREstate{}}& \gramdef & \STATE*{} \gramstar{\state*{}}                           \ACTIONS*{}
  \\& & \\

\gramfunc{\state{}}& \gramdef & \gramopt{\CONST*{}} \decl*{} \gramopt*{\EQ*{}
                                \expr*{}}
  \\& & \\

\gramfunc{\decl{}}& \gramdef & \gramsepnelist{\COMMA*{}}{\IDENT*{}}             \COLON*{} \ty*{}
\end{obeliskgrammar}

A state variable can optionally be declared as constant or with expressions that constrain the possible values of that variable. Comma separation can be utilized as a syntactic shorthand to declare multiple variables of the same type with minimum boilerplate. Besides declaring variables, another critical part of the state declaration is that it may define custom types to be used throughout the concept. 

\subsection{Types}
Alloy conforms to first-order logic, but its syntax may admit higher-order quantifications \cite{Alloy6Documentation}. Many such constraints can be reduced to first order during analysis, e.g. by using Skolemization, or produce warnings when this is impossible. \textit{Conceptual} makes no such guarantees and enforces first-order type. This is largely achieved with the parsing productions, but additional checks for certain expressions may be made during semantic analysis.

\begin{obeliskgrammar}
\gramfunc{\primUNDERSCOREty{}}& \gramdef & \gramhor{\STR*{}, \INT*{}, \IDENT*{}} \\& & \\

\gramfunc{\mult{}}& \gramdef & \gramhor{\SET*{}, \ONE*{}, \LONE*{}}
  \\& & \\
  
\gramfunc{\ty{}}& \gramdef & 
\gramhor{\gramopt{\mult*{}} \primUNDERSCOREty*{},\primUNDERSCOREty*{} \ARROW*{} \ty*{}}
  \\& & \\
\end{obeliskgrammar}

Notably, the parser does not include boolean types. This is for the same reason that Alloy\footnote{Alloy has a module implementing booleans, but the developers typically advise against using it, noting that boolean types are often the wrong way to do design modeling \cite{DanielJacksonSoftwareAbstractions}.} does not: having booleans implies having sets of booleans. Supposing that $T$ and $F$ denote true and false values respectively, then, as an example, it becomes problematic to interpret the meaning of expressions that evaluate to either $\{\}$ or $\{T,F\}$. To make this work, one could extend the interpretation of the logical operators over the sets of booleans, introducing additional complexity \cite{DanielJacksonSoftwareAbstractions}.

\subsection{Actions}
To fulfill its purpose, the concept must behave in a certain way. Actions define this behavior and have quite a few moving parts:

\begin{obeliskgrammar}

\gramfunc{\lval{}}& \gramdef & 
\gramhor{\IDENT*{},
 \lval*{} \DOT*{} \lval*{}}
  \\& & \\
    
\gramfunc{\assign{}}& \gramdef & 
\gramhor{
\COLON*{} \EQ*{},
\binop*{} \EQ*{}
}
  \\& & \\

\gramfunc{\stmt{}}& \gramdef & \gramsepnelist{\COMMA*{}}{\lval*{}} \assign*{}
                               \expr*{}
 \\& & \\

\gramfunc{\actionUNDERSCOREsig{}}& \gramdef & \ACT*{} \LPAR*{}
\gramseplist{\COMMA*{}}{\decl*{}}
\RPAR*{}
  \\& & \\

\gramfunc{\actionUNDERSCOREfiringUNDERSCOREcond{}}& \gramdef & \gramopt*{\WHEN*{}
\expr*{}}
  \\& & \\

\gramfunc{\action{}}& \gramdef & \actionUNDERSCOREsig*{}
                                 \actionUNDERSCOREfiringUNDERSCOREcond*{}
                                 \gramstar{\stmt*{}}\\
  & \grambar &\actionUNDERSCOREsig*{} \COLON*{} \ty*{} \expr*{}
  \\& & \\

\gramfunc{\cUNDERSCOREactions{}}& \gramdef & 
\ACTIONS*{}
\gramplus{\action*{}} \OP*{}
  \\& & \\
\end{obeliskgrammar}

Notably, the action signature uses a special token \kw{ACT}, which functionally is the same as an \kw{IDENT}. This token, however, is only produced during lexing when a left parenthesis immediately follows an identifier, e.g. \kw{foo(}. An additional parenthesis token is inserted using the \kw{TokenCache} module for readability in the production rule. In the above rules, it is evident that \textit{Conceptual} distinguishes between two types of action. It supports actions that query the state and return a value; these only take an expression. The other type of action consists of a list of statements for mutating the state and may only optionally fire based on a boolean condition. This is similar to the notion of a precondition but (unlike a precondition) ensures that the action cannot happen unless the condition holds. The mutations can either be simple assignments or use C-style shorthands, such that a statement of the form

\begin{myalign}
    $e1 \enspace op= \enspace e2 $ & stands for $e' \enspace = \enspace e1 \enspace op \enspace e2$ \\
\end{myalign}

with $e'$ denoting the result of the operation. As actions are largely comprised of mutations to the state, these shorthands are effective at succinctly representing changes to types such as sets and relations, not just integers. This is not valid for all operations, though. For example, a relational product will never be well-typed using this shorthand. Such cases are rejected before proceeding to semantic analysis. 

\subsection{Operational Principle}
To reiterate, operational principles are intended to describe archetypical scenarios of a concept. Given the principle's significance to the concept, it might appear odd that it can parsed harmlessly as a comma-separated list of expressions - especially since \textit{Conceptual} has no \textit{block} expressions.

\begin{obeliskgrammar}
\gramfunc{\cUNDERSCOREop{}}& \gramdef & \OP*{}
                                        \gramseplist{\COMMA*{}}{\expr*{}}
  \\& & \\
\end{obeliskgrammar}

The expressions must evaluate to booleans and each represents a certain concept property. To make this representation useful, though, \textit{Conceptual} introduces new boolean connectives for chaining expressions together. However, these connectives are exclusively for use within the operational principle.

\subsection{Expressions}
The expressions of \textit{Conceptual} generally fall into three categories: \textit{relational expressions}, \textit{boolean expressions}, and \textit{integer expressions}. The type of expression is not determined by the grammar but is inferred from the context by the type checker. Most operators apply only to one category of expression\footnote{Only a few operators are overloaded; for instance, '$+$' means different things when used on integers as opposed to relations.}. Generally, expressions are defined as: 

\begin{obeliskgrammar}
\gramfunc{\expr{}}& \gramdef & 
    \gramhor{\const*{}, 
    \unop*{} \expr*{},
    \lval*{},
    } \\
    & \grambar & \expr*{} \binop*{} \expr*{} \\
    & \grambar &\expr*{} \gramopt{\NOT*{}} \compareUNDERSCOREop*{} \expr*{}\\
   & \grambar &\expr*{} \LBRACK*{} \gramsepnelist{\COMMA*{}}{\expr*{}}
              \RBRACK*{}\\
  & \grambar &\LBRACE*{} \gramsepnelist{\COMMA*{}}{\decl*{}} \PIPE*{} \expr*{}
              \RBRACE*{}\\
  & \grambar &\gramopt*{\CAN*{} \gramopt{\NOT*{}}} \call*{}
  \\& & \\
\end{obeliskgrammar}

Of course, not all expressions can be used universally, and some operations require and/or produce specific types. This is touched on in more detail below.

\textbf{Relational} expressions have the following operations

\begin{obeliskgrammar}
    \gramfunc{\binop{}}& \gramdef & 
    \gramhor{
    \PLUS{},
    \MINUS{},
    \AMP{}
    \DOT{},
    \ARROW{},
    }
  \\& & \\

\gramfunc{\unop{}}& \gramdef & 
    \gramhor{
    \TILDE*{},
    \CARET*{},
    \STAR*{} \CARET*{},
    }
  \\& & \\
\end{obeliskgrammar}

With the following meanings

\begin{myalign}
    $\sim e$ & transpose of $e$; \\
    $^\wedge e$ & non-reflexive transitive closure of $e$; \\
    $*^\wedge e$ & reflexive transitive closure of $e$; \\
    $e1 \enspace+\enspace e2$ & union of $e1$ and $e2$; \\
    $e1 \enspace -\enspace e2$ & difference of $e1$ and $e2$; \\
    $e1 \enspace \& \enspace e2$ & intersection of $e1$ and $e2$; \\
    $e1 \enspace.\enspace e2$ & join of $e1$ and $e2$; \\
    $e1 \rightarrow e2$ & product of $e1$ and $e2$.\\
\end{myalign}

\textbf{Integer} expressions have these operations

\begin{obeliskgrammar}
    \gramfunc{\binop{}}& \gramdef & 
    \gramhor{
    \PLUS{},
    \MINUS{},
    \STAR{},
    \SLASH{},
    \PERCENT{},
    }
  \\& & \\

\gramfunc{\unop{}}& \gramdef & \#
  \\& & \\
\end{obeliskgrammar}

The expression $\#e$ denotes the cardinality of $e$. The rest have meanings from standard arithmetic.

Finally, despite not giving users access to booleans, \textbf{boolean} expressions still exist. Similar to Alloy, they include standard logical connectives, negation, equalities, inequalities, and operations to check for subsets (more on this in section~\ref{sec:semantic_type_checking}). In addition, \textit{Conceptual} introduces new connectives between expressions that are only allowed in the operational principle. One such operation is sequential composition\footnote{This also has a dual syntax, and \textit{a then b} or \textit{a;b} can be used interchangeably.}. An expression such as \textit{a;b} would then signify that after \textit{a} then \textit{b} holds\footnote{Both \textit{a} and \textit{b} must evaluate to booleans. However, calls to actions that mutate the state evaluate to a boolean. This is a consequence of the translation in section~\ref{sec:cg_translating_concepts}.} in the state immediately following after \textit{a}. Another new expression is \textit{until}, which signifies that the left expression holds until the right expression is true. Finally, \textit{Conceptual} also introduces two new unary operations for the operational principle. The first operation, \textit{can}, extracts the precondition of an action; thus, an expression \textit{can a} holds in a state where action \textit{a} can happen. The second is \textit{no}, which, in an expression \textit{no a}, evaluates to true if \textit{a} has never been true historically and it is false otherwise\footnote{This is useful for checking if a specific action has been run previously.}. 

\begin{obeliskgrammar}
    \gramfunc{\binop{}}& \gramdef & 
    \gramhor{
    \LAND*{},
    \LOR*{},
    \THEN*{},
    \UNTIL*{}
    } 
  \\& & \\

\gramfunc{\compareUNDERSCOREop{}}& \gramdef & 
\gramhor{
\EQ*{},
\IN*{},
\LT*{},
\GT*{},
\LTE*{},
\GTE*{}
}
  \\& & \\

\gramfunc{\unop}& \gramdef & 
\gramhor{
\NOT*{},
\NO*{}
}
  \\& & \\
\end{obeliskgrammar}

Notice in the general definition of the expression that comparison operators can be negated. An added benefit of parsing it this way is that a token for "!=" is redundant. In fact, by doing this with two separate tokens, one benefits both from the dual syntax and can also include whitespace between the tokens.

\subsection{Apps}
Apps consist of a list of concept instantiations or dependencies and a way of composing those concepts. Presently, apps compose concepts purely via synchronizations of their actions. Consequently, in terms of parsing, apps consist largely of glorified calls with an attached namespace pointing to the concept with that action. A synchronization distinguishes between the first call (the trigger action) and the subsequent calls (response actions)\footnote{This is elaborated on in section~\ref{sec:semantic_anaylsis_traversal} and chapter~\ref{chap:cg_code_generation}.}. 

\begin{obeliskgrammar}
    \gramfunc{\appUNDERSCOREdep{}}& \gramdef & \filepath*{} \gramopt*{\LBRACK*{}
                                           \RBRACK*{}}\\
  & \grambar &\filepath*{} \LBRACK*{}
              \gramsepnelist*{\COMMA*{}}{\gramopt*{\IDENT*{} \DOT*{}}
              \primUNDERSCOREty*{}} \RBRACK*{}
  \\& & \\

  \gramfunc{\trigger{}}& \gramdef & \IDENT*{} \DOT*{} \multUNDERSCOREcall*{}
  \\& & \\

\gramfunc{\response{}}& \gramdef & \IDENT*{} \DOT*{} \call*{}
  \\& & \\

\gramfunc{\sync{}}& \gramdef & \sync*{} \trigger*{} \gramplus{\response*{}}
\\& & \\

\gramfunc{\app{}}& \gramdef & \APP*{} \IDENT*{} \INCLUDE*{}
                              \gramplus{\appUNDERSCOREdep*{}}
                              \gramstar{\sync*{}}
  \\& & \\

\end{obeliskgrammar}

A \kw{mult\_call} is essentially just a regular call; the difference is that variable arguments may be prefixed with a multiplicity to denote that the synchronization does not occur each time the action is executed - but only on select inputs. The complexity of apps, though, largely revolves around keeping track of the current and valid namespaces, primarily affecting semantic analysis (chapter~\ref{chap:semantic_analysis}) and code generation (chapter~\ref{chap:cg_code_generation}). 

\subsection{Precedence and Associativity}
The precedence of expression operators is listed below, the tightest first: 

\begin{itemize}
    \item unary operators: \kw{TILDE}, \kw{CARET}, and \kw{STAR CARET};
    \item dot join: \kw{DOT};
    \item box join: \kw{LBRACK} \kw{RBRACK};
    \item relational product: \kw{ARROW};
    \item binary operators: \kw{AMP}, \kw{STAR}, \kw{SLASH} and \kw{PERCENT};
    \item cardinality: \kw{CARD};
    \item binary operators: \kw{PLUS} and \kw{MINUS};
    \item negation of comparisons: \kw{NOT};
    \item comparison operators: \kw{IN}, \kw{EQ}, \kw{LT}, \kw{GT}, \kw{LTE}, and \kw{GTE}.
\end{itemize}

Then comes a bunch of logical operators with even lower precedence, again with the tightest first:

\begin{itemize}
    \item negation: \kw{NOT};
    \item conjunction: \kw{LAND};
    \item disjunction: \kw{LOR};
    \item unary \gls{ltl} operator: \kw{NO};
    \item binary \gls{ltl} connective: \kw{UNTIL};
    \item binary \gls{ltl} connective: \kw{THEN}.
\end{itemize}

All binary operators generally associate to the left. The sole exception is the \kw{THEN} operation, which associates to the right. Comparison operators are non-associative.  

\section{Testing the Parser}\label{parsing:sec_testing}
\begin{figure}[ht!]
    \centering
    \includegraphics[width=0.9\linewidth]{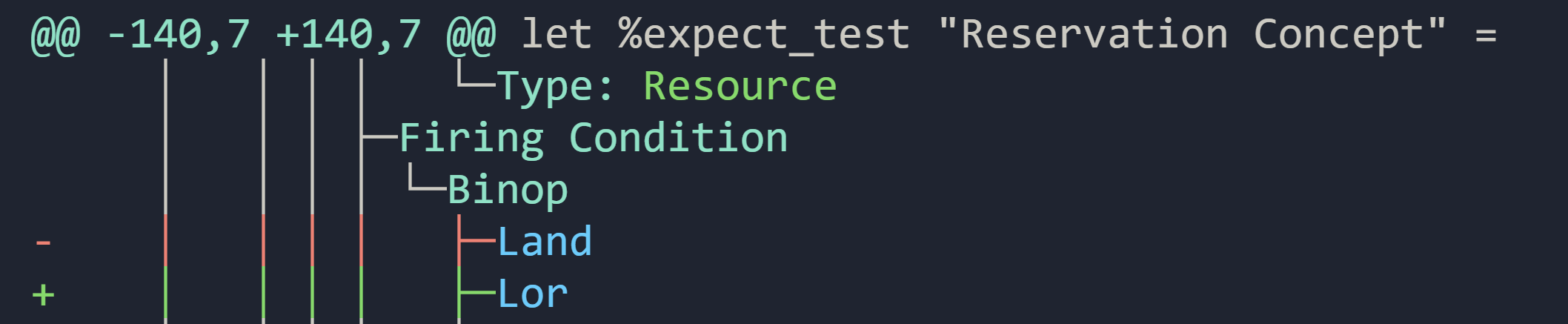}
    \caption[Example Unit Test]{Example of a unit test where a node in the \gls{ast} was changed from a conjunction to a disjunction to simulate a fault in the parser. The test catches the misbehavior and highlights the discrepancies from the expected parse tree.}
    \label{fig:unit-test-example-parsing}
\end{figure}

A suite of unit or \textit{expect} tests are used to verify the correctness of the syntax implementation. Each test takes a particular model and examines whether the pretty-printed \gls{ast} from parsing conforms to the expected structure. As shown in figure~\ref{fig:unit-test-example-parsing}, these tests are designed to catch faulty behavior in the parser, such as incorrect precedence or representation of operations in the \gls{ast}. The test corpus largely comprises the cases explored in chapter~\ref{chap:cases}.

\chapter{Semantic Analysis in Conceptual}\label{chap:semantic_analysis}
\textit{This chapter explores various aspects of how \textit{Conceptual} analyzes the semantics of the generated \gls{ast}.
First, section~\ref{sec:semant_semantics} stresses the lack of a complete account of the semantics. Following this, section~\ref{sec:semantic_symbol_table_scope} briefly accentuates the need for a symbol table and describes how scopes are handled. Then, section~\ref{sec:semantic_namespaces} outlines the different namespaces in \textit{Conceptual} and section~\ref{sec:semantic_error_reporting} how errors are reported. Subsequently, section~\ref{sec:semantic_type_checking} discusses some aspects of the type system. Afterward, section~\ref{sec:semantic_anaylsis_traversal} provides general comments on the traversal of the \gls{ast} and the analysis process. Finally, section~\ref{semant:sec-testing} touches on testing of the semantic analysis phase.} 

\includegraphics[width=\textwidth]{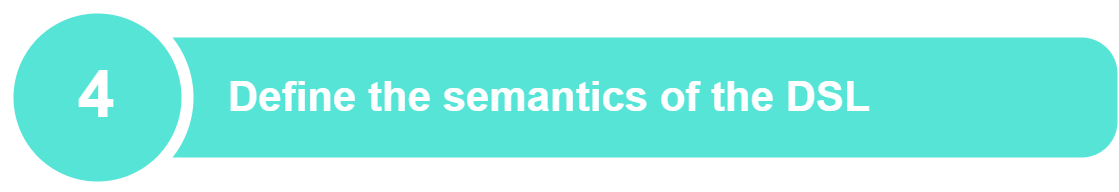}
\includegraphics[width=\textwidth]{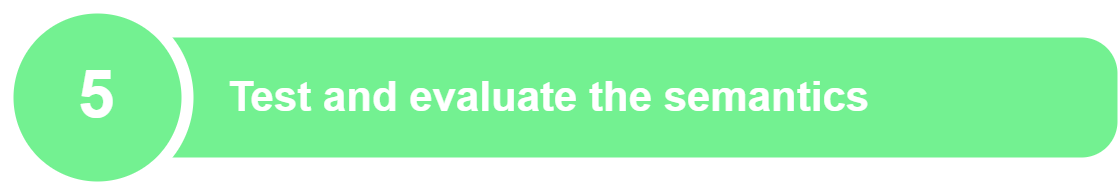}

\section{Semantics}\label{sec:semant_semantics}
It should be emphasized that the thesis does not explicitly define the complete semantics of \textit{Conceptual} as it does with the syntax. Although the meaning of appropriate language constructs may have been described informally until this point, especially when they differ from Alloy, the coverage is incomplete and is spread sporadically across different sections of the thesis. The decision to omit a detailed account is primarily influenced by \textit{Conceptual}'s close resemblance to Alloy's syntactic and semantic structures. Additionally, certain language features, such as boolean literals or the ability to generate new variables, may have been excluded preemptively due to the challenge of modeling them in Alloy. Thus, it is probable that \textit{Conceptual} does not adequately capture the desired semantics of a concept-oriented \gls{dsl}. Ultimately, while the thesis is more concise on account of this omission, the semantics must be implicitly inferred based on the translation to Alloy as a detriment. 

\section{Symbol Table and Scope}\label{sec:semantic_symbol_table_scope}
\gls{dsl}s generally tend to exclude variables, loops, and conditionals \cite{FowlerDSL}. This is largely true for \textit{Conceptual}, which does not contain imperative control structures. There are no procedure calls, and actions are defined independently of others. However, variables exist in \textit{Conceptual} but can only be declared in the state component. A symbol table of variable- and action declarations is maintained to ensure all used variables are explicitly declared and only identify a single object at a time. Concretely, \textit{Conceptual} implements the symbol table using a map and stores it in an environment. This environment is propagated during semantic analysis and represents the active scope. Scope resolution is resolved simply by not returning the environment after a function call. The notion of scope is very limited in \textit{Conceptual}: Each concept has a unique scope containing the state variables and the action signatures. Each action within a concept defines a nested scope that additionally includes action parameters. Concepts are defined in a scope that is accessible by apps. 

\section{Namespaces}\label{sec:semantic_namespaces}
There are four namespaces in \textit{Conceptual}:

\begin{itemize}
    \item custom types;
    \item variables, parameters, and action signatures;
    \item concepts;
    \item apps. 
\end{itemize}

Each identifier belongs to a single namespace, and identifiers in different namespaces may share names without risk of name conflicts. Within the same namespace, the same name may not be reused, and all forms of shadowing are disallowed. 

\section{Error Reporting}\label{sec:semantic_error_reporting}
Encountered errors are accumulated in the environment as they are detected. This allows all errors to be reported collectively at the end of semantic analysis as opposed to sequentially reporting the first error. A custom \kw{Error} module defines the different types of semantic errors and associates a message with each, relying on the \kw{Location} module to format the line and character numbers stored in each \gls{ast} node. Of course, storing the errors in the environment presents a challenge: when a scope is resolved, newly accumulated errors will effectively be forgotten. \textit{Conceptual} achieves error persistency across scopes by maintaining the list of errors in the environment via reference, violating immutability in this single case. 

\section{Type Checking}\label{sec:semantic_type_checking}
Unlike the Alloy type system \cite{EdwardsJonathan_Alloy_Typesystem}, which can largely be regarded as untyped as type errors are only produced when expressions (other than the constant \kw{none}) can be shown to reduce to an irrelevant expression\footnote{Replaceable by the empty relation without changing the meaning of the affected formula.}, \textit{Conceptual} behaves similarly to strongly-typed languages. Although the lack of runtime errors makes type soundness less critical in \gls{mde}, being able to make stronger guarantees based on type information can hardly be called detrimental. Such guarantees come at the cost of flexibility - though the cost is mitigated due to the lack of subtyping in \textit{Conceptual}.  

Additionally, a large aspect of semantic analysis involves ensuring a level of type safety. Beyond the primitive types\footnote{Strings, integers, and booleans}, each concept commonly defines a set of custom types and/or is parameterized by generic types in the concept header. Custom-type declarations are only deemed when made within the concept header or its state declaration. Valid custom types are likewise stored in the environment and propagated through it. A drawback of Alloy is its limitation to first-order logic, necessitating that semantic analysis must verify that only first-order types are constructed and ruling out sets of sets or relations over sets. As \textit{conceptual} distinguishes between types, sets of types, and relational types, the type system utilizes a degree of type coercion and automatically converts types where appropriate. This is particularly useful for expressions such as a union of a set with a particular element. 

Types can generally be inferred from expressions, e.g. types of variables and actions through an environment lookup or based on the operator. Binary operations like relational join must ensure that the subexpressions are compatible to produce a valid composite type. Similarly, the \textit{in} operation requires a compatibility check between the left and right subexpressions. Interestingly, \textit{Conceptual} allows the "in" operator to be used with simple types and relations. Consider an element $e$ that belongs to a set $E$ (denoted as $e \in E$), and a relation $\mathcal{R}$ that encompasses $E$ among its sets, represented as $\mathcal{R} \subseteq A \times B \times ... \times E \times ...$. In this context, the operation $e\in \mathcal{R}$ checks whether the element $e$ is present in any set $E$ in the relation $\mathcal{R}$. This is used by the \kw{retract} action in the reservation concept (listing~\ref{lst:concept_reservation}).

\textit{Conceptual} uses a form of bidirectional typing relying largely on two functions: \kw{infertype_expr} (line one) and \kw{typecheck_expr} (line two) with the signatures seen in listing~\ref{lst:bidirectional_typing_signatures}. 

\begin{lstlisting}[caption={[Signatures for Bidirectional Typing] Signatures for type checking and type inference. }, float=ht, label={lst:bidirectional_typing_signatures}]
Env.environment -> Ast.expr -> TypedAST.expr * TypedAST.typ
Env.environment -> Ast.expr -> TypedAst.typ -> TypedAST.expr
\end{lstlisting}

The former synthesizes or deduces the type when no explicit type is given. On the contrary, the latter checks that the expression has an appropriate type. Both return an explicitly typed expression. Such a type system can use checking to support features for which inference is undecidable - and inference to avoid the large annotation burden \cite{Bidirectional_typechecking_Dunfield_2021}.

The typed \gls{ast} includes a new error type for cases when the type cannot be inferred. The type checker treats the error type as universal and equivalent to all other types. This essentially makes expressions that include the error type agnostic regarding types, which helps prevent errors from propagating across different expressions. 

\section{Analysis Process}\label{sec:semantic_anaylsis_traversal}
Generally, \textit{Conceptual} only traverses the \gls{ast} once, with the sole exceptions being the declaration of concept states and the inclusion of dependencies. In these cases, state variables may be defined in terms of other variables; app dependencies may be parameterized over other concepts. To allow mutually recursive definitions and/or for them to be declared in any arbitrary order, an initial pass must be performed over the nodes corresponding to the state and the app dependencies. This pass adds the necessary information to the environment before fully processing the nodes. Contrary to the concept state, a single pass suffices for actions that are defined independently of each other. 

In analyzing the \gls{ast}, concepts are first processed individually with a fresh\footnote{Not accounting for the errors or general concept symbol table.} environment. After analyzing a particular concept, its environment — consisting of information such as the available actions, variables, and types — is saved to a nested symbol table within the main symbol table. When apps are analyzed, it is possible to switch the context to a particular concept as needed. Subsequently, the apps are analyzed. During this analysis, the availability of the included concepts is first checked. If a concept is missing from the current file, the application will attempt to dynamically include it through the file path in the include-clause\footnote{The \kw{.con} file extension is automatically added if omitted.}. The file is compiled to the \gls{ast} level if found. It then searches within this \gls{ast} for the missing concept specified by the basename of the file path. If the missing concept is found, it combines the \gls{ast} of the particular concept with its own \gls{ast}; otherwise, the specification is considered erroneous. 

Furthermore, while nodes are generally analyzed locally, it is worth noting that the environment maintains flags to signify whether the current context is in the operational principle, in synchronizations, or elsewhere. This is necessary as a few expressions, such as calls and the \gls{ltl} operators, are only allowed within certain contexts. Another detail is that, within those contexts, instead of resulting in an error, undeclared variable arguments to call are inserted into the environment as new temporary variables. These variables are only \textit{alive} for a single operational principle or synchronization.

\section{Testing Semantic Analysis}\label{semant:sec-testing}
Almost identical to the approach for testing the parser, the correctness and completeness of the semantic analysis phase is assessed by examining the generated \gls{ast}. Similarly, the test corpus remains the same but with the addition of several negative control tests. The negative controls aim to catch certain semantic inconsistencies, such as using undeclared variables, by providing buggy models and verifying that the produced environment is non-empty and populated with error(s).

\chapter{Code Generation}\label{chap:cg_code_generation}
\textit{This section describes the last phase of the proposed compiler: code generation. Section~\ref{sec:cg_workings_of_code_generator} first outlines the overall behavior of the code generator. Then section~\ref{sec:cg_translating_concepts} discusses how concepts are translated to Alloy. Afterward, section~\ref{sec:cg_translating_apps} details the translation of apps. Finally, section~\ref{cg:sec-testing} discusses how code generation was tested. }

\includegraphics[width=\textwidth]{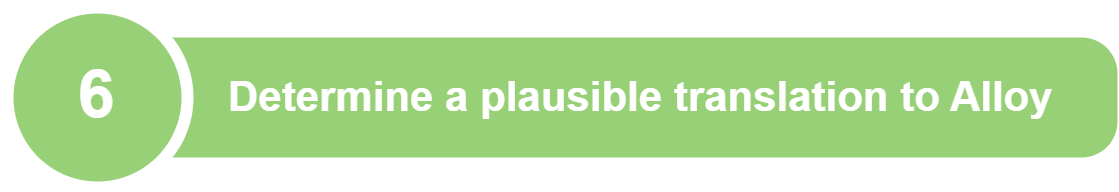}
\includegraphics[width=\textwidth]{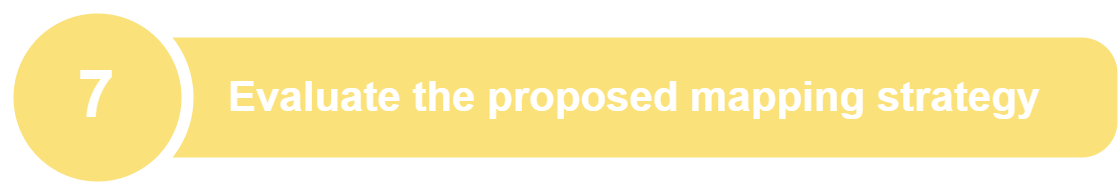}

\section{General Workings of Code Generator}\label{sec:cg_workings_of_code_generator}
The typed \gls{ast} is traversed in a way similar to how the \gls{ast} was traversed in section\ref{sec:semantic_anaylsis_traversal}: mostly with a single-pass, maintaining context in an environment, and nodes are processed locally. Nodes for concepts and apps are each translated into an Alloy-\gls{ast}-esque \gls{ir}, which is then serialized into a separate file.

\section{Translating Concepts}\label{sec:cg_translating_concepts}
Since concepts define self-contained state machines, their translations should ideally reflect this and be freestanding and independent. Consequently, concepts are translated into Alloy \textbf{modules}. Moreover, Alloy conveniently supports generic modules parameterized over certain signatures, effectively describing abstract concepts. 

\subsection{Purpose}
A concept's purpose, functionally, does not affect the behavior of the state machine. Thus, they \textit{could} be omitted entirely from the generated code. Nevertheless, \textit{Conceptual} embeds the purpose into the module as an Alloy \textbf{comment}. 

\subsection{State}
To effectively model state machines, the state must be able to \textit{change}. This suggests the possibility of using Alloy6's variable signatures and fields, which is the approach adopted by \textit{Conceptual}. Specifically, states are translated into \textbf{mutable fields} enclosed within a special state signature declared with a multiplicity of one, i.e.\ a concept-global singleton state. This mapping is the most direct, bearing the closest resemblance to the input. Additionally, concept states may define custom types. New types are translated into empty, \textbf{top-level signatures}, inherently disjoint in Alloy. States may also be constrained by a particular expression. Such constraints are mapped to separate \kw{fact} entities. 

Alloy, however, has no special syntax for declaring the transitions system\footnote{Instead, one must specify the transition system implicitly by means of an \gls{ltl} formula that recognizes the valid execution traces. An idea originally presented by Leslie Lamport \cite{LamportTemporalLogicOfActions}.}.
Specifying a formula that defines the initial valid states is generally required for dynamic models. Without such a constraint, models may include traces starting in any state - even ones that \textit{should} not be allowed by the model. \textit{Conceptual} imposes the constraint that all mutable states are empty initially. This constraint is realized with a \kw{fact} containing no temporal operators\footnote{Formulas without \gls{ltl} operators are only required to hold in the initial state of every trace.}.

\subsection{Actions}
As mentioned in chapter~\ref{chap:parsing_conceptual}, \textit{Conceptual} distinguishes between two types of actions: those that query the state and those that mutate it. These are translated differently. The former is translated into \textbf{functions} and the latter into \textbf{predicates}. Each predicate effectively evaluates to its firing condition. Additionally, in both cases, a separate predicate is created. This predicate has the same name as the action but prefixed with \textit{\_can\_}; its body consists of the action's firing condition\footnote{In Alloy empty predicates evaluate to true. Therefore, the new predicate effectively tells if the action can happen.}, if any. 

To complete the transition system of the state machine, the model should specify which transitions are possible at each state. In \textit{Conceptual}, the set of transitions is constituted by actions that mutate the state and a special ``do nothing" transition. Given that the firing condition of an action already acts as a guard for determining if an action can happen, the set of transitions can be defined to be valid in each state using the \kw{always} operator. An important detail regarding mutable events is that one must consider their effects on ALL fields comprising the state. The state can freely evolve if nothing is specified and the event occurs. Consider the translation in figure~\ref{fig:frame_conditions}. If line eight of the Alloy snippet was omitted, \kw{reservations} could take any value not prohibited by external formulas after running \kw{provide}. 

\begin{figure}[ht!]
    \centering
    \begin{adjustwidth}{-2.15cm}{-2.15cm} 
    \begin{minipage}{0.56\textwidth}
\begin{lstlisting}[style=concept, showlines=true] 
state 
  available: set Resource
  reservations: User -> set Resource
actions
  provide(r: Resource)
    available += r 



\end{lstlisting}
    \end{minipage}\hfill
    \begin{minipage}{0.7\textwidth}
        \begin{lstlisting}[style=alloy, numbers=right]
one sig State { 
  var available : set Resource,
  var reservations : User -> set Resource 
}

pred provide[r : Resource] {
  (State.available') = (State.available) + r
  (State.reservations') = (State.reservations)
}
\end{lstlisting}
    \end{minipage}
        \end{adjustwidth}
        \caption[Example of a Frame Condition]{Example to explain frame conditions. The left snippet is part of the reservation concept in \cite{JacksonConcepts}. The right snippet is a translation to Alloy. The \kw{'} operator refers to the value of a mutable expression in the subsequent state. It may be nested.}
        \label{fig:frame_conditions}
\end{figure}

Commonly, the intention is for these fields to remain unchanged. However, this must be explicitly declared. A ``no change" effect is often called a frame condition. Consequently, the body of actions that mutate the state is generally structured in the following way: the firing condition, then the intended effect of that action, and finally frame conditions preserving the previous state of the remaining fields. It is worth noting that the \kw{'} operator increments the internal clock in Alloy once. This is important to keep in mind when translating actions of multiple statements. In particular, if one wants to use \kw{always} to check assertions across all possible states, actions should ideally only increment the clock a single time internally. The reason is that, in a sense, the "internal clock" is local to each particular temporal formula. To showcase this, listing~\ref{lst:provide_v2} depicts another provide predicate that adds two resources by nesting \kw{'}.

\begin{lstlisting}[style=alloy,caption={[Alternative Provide Predicate]Alternative provide predicate for adding two resources.}, label={lst:provide_v2},float=ht]
pred provide2[r1,r2 : Resource] {
  (State.available') = (State.available) + r1
  (State.available'') = (State.available') + r2
  (State.reservations') = (State.reservations)
  (State.reservations'') = (State.reservations')
}
\end{lstlisting} 

The \textit{issue} then is that assertions of the form in listing~\ref{lst:provide_v2_assertion} will not hold. Technically, the \kw{always} operator will test the assertion at every state $i \geq 0 $, and at every $i$ the antecedent is evaluated at $i$ and the consequent at $i+1$ due to \kw{after}. Thus, even if the antecedent internally inspects many future states, such as \kw{State.available''}, the consequent will still consider state $i + 1$. That is, the assertion is violated as $i+1$ is before \kw{r2} is included in the set of available resources. In other words, the formula cannot be verified as it does not hold in all \textit{intermediate} states. 

\begin{lstlisting}[style=alloy,caption={[Assertion for Provide Behavior]Asserting the intended behavior of \kw{provide2}.}, label={lst:provide_v2_assertion},float=ht]
assert ... { 
    always ( all r1, r2 : Resource | provide2[r1,r2] => 
        after r1 + r2 in State.available 
)}
\end{lstlisting}

To make the analysis more intuitive (allowing assertions such as listing~\ref{lst:provide_v2_assertion}) and the generated code more readable, the action is instead translated as seen in figure~\ref{fig:accumulative_action} by traversing all statements and accumulating expressions modifying the same variable. Actions, however, are intended to be read in a purely declarative manner, where the order of the statements is irrelevant. Naturally, this raises the question of how \textit{Conceptual} handles actions that must accumulate compound expressions of different precedence (e.g. set union and intersection) or consist of both compound and non-compound assignments. Rather than disallowing the former, the statements are interpreted in the order in which they are written. The latter is prohibited. 

\begin{figure}[ht!]
    \centering
    \begin{adjustwidth}{-2.15cm}{-2.15cm} 
    \begin{minipage}{0.52\textwidth}
\begin{lstlisting}[style=concept, showlines=true] 
provide2(r1 , r2: Resource)
    available += r1 
    available += r2
    
\end{lstlisting}
    \end{minipage}\hfill
    \begin{minipage}{0.75\textwidth}
        \begin{lstlisting}[style=alloy,numbers=right]
pred provide[r : Resource] {
  (State.available') = (State.available) + r1 + r2
  (State.reservations') = (State.reservations)
}
\end{lstlisting}
    \end{minipage}
        \end{adjustwidth}
        \caption[Accumulative Translation of Action]{Translation of \kw{provide2} from \textit{Conceptual} (left) to Alloy (right).}
        \label{fig:accumulative_action}
\end{figure}

\subsection{Operational Principle}
In the operational principle, properties of archetypical scenarios are expressed. These intended properties, however, may not logically follow from a given concept specification (as will be shown in section~\ref{subsec:concept_label}). Thus, it is imperative to either prove one's assumptions about the model or recognize the model as flawed. \textit{Conceptual} translates each scenario within the operational principle to an \textbf{assertion}, effectively off-loading the burden of proof to Alloy, which can exhaustively search for counterexamples within a specified scope. Each assertion uses the \kw{after} operator, implying that the property must be true in all states. 

\section{Translating Apps}\label{sec:cg_translating_apps}
Like concepts, each app is mapped to a unique Alloy file. External concepts are then included simply by opening their respective modules, specializing any abstract concepts in the process. Synchronizations are translated into implication formulas using the \kw{always} operator, all contained within a \kw{fact} clause. The antecedent of the implication is the trigger action; the consequent is a conjunction of all response actions. Semantically, synchronizations are transactional and are assumed not to happen when a response action is impossible: either every action occurs or none do.  

\section{Testing Translation}\label{cg:sec-testing}
As was the case for section~\ref{parsing:sec_testing} and \ref{semant:sec-testing}, the testing of the code generation phase is also partly \gls{ast}-based, verifying whether input models are correctly translated into their corresponding Alloy \gls{ir}. Moreover, the concrete serialization to Alloy files is also tested. Notably, these tests only assert that the translation aligns with expectations and do not necessarily guarantee that the generated model runs seamlessly. Automating a testing framework for such a task is complicated because the most straightforward approach involves integration with the Alloy \gls{api}. Again, the test corpus largely comprises the cases in chapter~\ref{chap:cases}, for which the desired behavior has been verified manually.

\chapter{Case Studies}\label{chap:cases}
\textit{This chapter models and runs the proposed compiler on several valid concepts and apps expressed in Conceptual. Due to its verbosity from boilerplate code, the generated Alloy code is not generally included in this chapter. Structurally, section~\ref{sec:case_concepts} and its subsections (\ref{subsec:concept_todo} through \ref{subsec_concept_email}) present a variety of concepts. On the other hand, section~\ref{sec:case_concept_composition} includes two examples (\ref{subsec:todo-label-app} and \ref{case:subsec_todo_label_email_app}) of how concepts can be composed using synchronizations.}

\includegraphics[width=\textwidth]{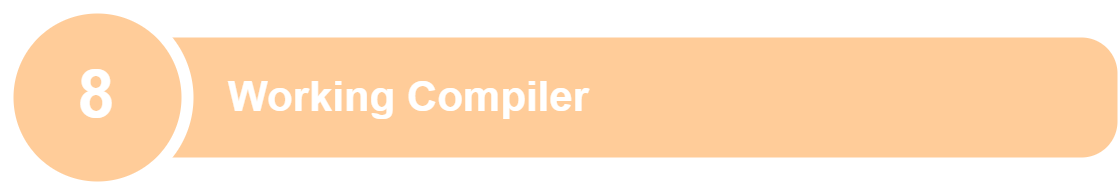}

\section{Concepts}\label{sec:case_concepts}
This section depicts several concepts, from \cite{JacksonConcepts} unless otherwise explicitly stated, all expressed in \textit{Conceptual}. These concepts are generally presented with minimal context, aiming to demonstrate a few working examples and their concrete syntax. A few exceptions to this are section~\ref{subsec:concept_reservation} and \ref{subsec:concept_label}, which each locates errors in the concept-specification from \cite{JacksonConcepts} and proposes how they can be fixed. 

\newpage
\subsection{Todo}\label{subsec:concept_todo}
The todo concept partitions a set of tasks into two sets, \kw{done} and \kw{pending}. Tasks are marked as \textit{pending} when they are initially added and transition to \textit{done} upon \textit{completion}.

\begin{lstlisting}[style=concept, float=ht,caption={[Todo Concept]The todo concept defined.},label={lst:concept_todo}]
concept todo 
purpose "keep track of tasks"
state done, pending: set Task
actions
  add(t : Task)
    when t not in done+pending
        pending += t
  delete(t : Task)
    when t in done+pending 
        done -= t
        pending -= t
  complete(t : Task)
    when t in pending 
        pending -= t
        done += t
principle
  add(t) then t in pending until delete(t) or complete(t),
  complete(t) then t in done until delete(t)
\end{lstlisting}

\subsection{Style}\label{subsec:concept_style}
The style concept associates elements with a particular style, automatically reformating the elements if the style is redefined.
\begin{lstlisting}[style=concept,float=ht,caption={[Style Concept]The style concept defined.},label={lst:concept_style}]
concept style [Element, Format]
purpose "easing consistent formatting of elements"
state
  assigned: Element -> one Style 
  defined: Style -> one Format
  format: Element -> one Format = assigned.defined
actions
  assign(e:Element, s:Style)
    e.assigned := s
  define(s:Style, f:Format)
    s.defined := f
principle
  define(s,f); assign(e1,s); assign(e2,s); define(s,f2) then
        e1.format = f2 and e2.format = f2
\end{lstlisting}

\newpage
\subsection{Trash}\label{subsec:concept_trash}
The trash concept maintains two sets of items: one where items are fully accessible and one for temporary storage of items that have been \textit{deleted}.  

\begin{lstlisting}[style=concept, float=ht,caption={[Trash Concept]The trash concept defined.},label={lst:concept_trash}]
concept trash [Item]
purpose "to allow undoing of deletions"
state accessible, trashed : set Item
actions
  create(x : Item)
    when x not in accessible + trashed
      accessible += x
  delete(x : Item)
    when x in accessible and x not in trashed
      accessible -= x
      trashed += x
  restore(x : Item)
    when x in trashed
      trashed -= x
      accessible += x
  clear()
    when trashed != empty
      trashed := {}
principle 
  delete(x); restore(x); x in accessible,
  delete(x); clear(); x not in accessible + trashed
\end{lstlisting}

\newpage
\subsection{Upvote}\label{subsec:concept_upvote}
The upvote concept defines a way to gauge user sentiment by letting users express their approval (\kw{upvote}) or disapproval (\kw{downvote}) of certain items, ensuring that users can only vote once per item. This concept does not originate from \cite{JacksonConcepts} but is outlined on the book's official site under case studies \cite{jackson2024essence_website}.

\begin{lstlisting}[style=concept,float=ht,caption={[Upvote Concept]The upvote concept defined.}, label={lst:concept_upvote}]
concept upvote [Item, User]
purpose "gauge user sentiment"
state upvotes, downvotes : Item -> set User
actions 
  upvote(i : Item, u : User)
    when u not in i.upvotes
      i.upvotes += u
      i.downvotes -= u 
  downvote(i : Item, u : User)
    when u not in i.downvotes
      i.downvotes += u
      i.upvotes -= u
  unvote(i : Item, u : User) 
    when u in i.(upvotes+downvotes)
      i.upvotes -= u 
      i.downvotes -= u
  count(i : Item) : int
    #i.upvotes - #i.downvotes
principle 
  upvote(i,u) or downvote(i,u) then can unvote(i,u),
  upvote(i,u)     then can not upvote(i,u) 
                              until unvote(i,u) or downvote(i,u),
  downvote(i,u) then can not downvote(i,u) 
                              until unvote(i,u) or upvote(i,u)
\end{lstlisting}

\newpage
\begin{lstlisting}[style=concept, float=ht, caption={[Reservation Concept]The reservation concept defined. Note that the specification in \cite{JacksonConcepts} is ambiguous; it is unclear if using a resource also consumes it. For a restaurant reservation, as used in the book, this would realistically be the case. However, reservations for books at a library or cloud computing resources permit \textit{indefinite} usage. This specification uses the latter design.},label={lst:concept_reservation}]
concept reservation [User, Resource]
purpose "manage efficient use of resources"
state 
  available: set Resource
  reservations: User -> set Resource
actions
  provide(r : Resource)
    available += r 
  retract(r : Resource)
    when r in available and r not in reservations 
      available -= r
  reserve(u : User, r : Resource)
    when r in available 
      u.reservations += r
      available -= r
  cancel(u : User, r : Resource) 
    when r in u.reservations 
      u.reservations -= r
      available += r
  use(u: User, r : Resource)
    when r in u.reservations
principle reserve(u,r) then can use(u,r) until cancel(u,r) 
\end{lstlisting}
\subsection{Reservation}\label{subsec:concept_reservation}
The reservation concept (see listing~\ref{lst:concept_reservation}) attempts to make efficient use of a pool of limited resources. Functionally, it uses a set to keep track of available resources, and its actions can associate these resources with specific users. Despite the operational principle holding universally, the specification likely behaves in a way that is not intended. Specifically, one can write an assertion, checking a formula such as listing~\ref{lst:formula_dup_reservation}, and discover that users can reserve the same resource.
\begin{lstlisting}[style=alloy,float=ht, caption={[Reservation Concept: Temporal Formula to Detect Design Flaw]A temporal formula that evaluates to true when a resource is unable to be both reserved and available simultaneously, false otherwise.},label={lst:formula_dup_reservation}]
always (
    no r : Resource, u : User | 
        r in State.reservations[u] and r in State.available
)
\end{lstlisting}
\begin{lstlisting}[style=concept, float=ht, caption={[Corrected Provide Action of Reservation Concept]Corrected version of the provide action in the reservation concept.}, label={lst:correction_provide}]
provide(r : Resource)
    when r not in reservations
        available += r 
\end{lstlisting}

\begin{figure}[ht]
    \centering
    \includegraphics[scale=0.75]{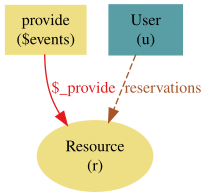}
    \caption[Reservation Concept Counterexample]{A counterexample found when checking the formula in listing~\ref{lst:formula_dup_reservation}. The entire trace is not included - only the state demonstrating the problematic behavior: if a user $u$ has a reservation $r$, $r$ can still be added to the available set through \kw{provide}. Ultimately, such behavior enables users to reserve the same resource.}
    \label{fig:reservation_concept_counterexample}
\end{figure}

Figure~\ref{fig:reservation_concept_counterexample} depicts a counterexample found by the Alloy analyzer, where resources are added to the set of available resources despite being reserved already. In other words, the concept as outlined in \cite{JacksonConcepts} unintentionally assumes that each resource provided is fresh. Consequently, the model effectively allows any number of users to reserve the same resource simultaneously, defeating the purpose of the concept. A possible solution, depicted in listing~\ref{lst:correction_provide}, is to include a firing condition that guards against this. If one wants to eliminate redundancy, adding the guard condition to \kw{provide} also makes it possible to refactor the condition for \kw{retract}. As resources can no longer be reserved and appear in the pool of available resources, the conjunction can be removed to better mirror the action for providing resources (see listing~\ref{lst:refactored_retract}). In fact, the firing condition could be removed entirely without adversely impacting the desired behavior. 

\begin{lstlisting}[style=concept, float=ht, caption={[Refactored Retract Action of Reservation Concept]Refactored version of the retract action after modifications to \kw{provide}.}, label={lst:refactored_retract}]
retract(r : Resource)
    when r in available
        available -= r
\end{lstlisting}

This example shows that coming up with an effective model is a non-trivial task, and even published and peer-reviewed literature such as \cite{JacksonConcepts} is not immune to design flaws. In addition, the value of incorporating an exhaustive analysis tool, such as a bounded model finder, is made evident. As demonstrated here, such tools frequently discover unintended behaviors, highlighting inadequacies or the omission of assumptions.

\newpage
\begin{lstlisting}[style=concept,float=ht,
caption={[Label Concept] The label concept defined.}, label={lst:concept_label}]
concept label [Item]
purpose "organize items into overlapping categories"
state labels: Item -> set Label 
actions 
  affix(i: Item, l : Label)
    i.labels += l
  detach(i : Item, l : Label)
    i.labels -= l
  find(l: Label) : Item 
    l.~labels
  clear(i : Item)
    i.labels := {}
principle
  affix(i,l) then i in find(l) until detach(i,l),
  no affix(i,l) or detach(i,l) then i not in find(l)
\end{lstlisting}

\subsection{Label}\label{subsec:concept_label}
\begin{figure}[ht!]
  \centering
  \begin{adjustwidth}{-1.5cm}{-1.5cm} 
    \begin{minipage}[b]{0.20\linewidth}
\includegraphics[width=0.9\linewidth]{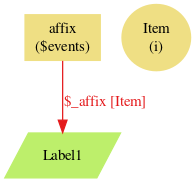}
      \caption*{(0)}
    \end{minipage}
    \hfill 
    \begin{minipage}[b]{0.20\linewidth}
\includegraphics[width=0.9\linewidth]{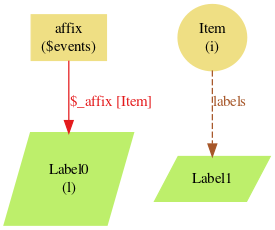}
      \caption*{(1)}
    \end{minipage}
    \hfill
    \begin{minipage}[b]{0.18\linewidth}
\includegraphics[width=0.9\linewidth]{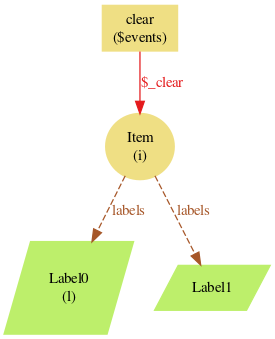}
      \caption*{(2)}
    \end{minipage}
    \hfill
    \begin{minipage}[b]{0.3\linewidth}
\includegraphics[width=\linewidth]{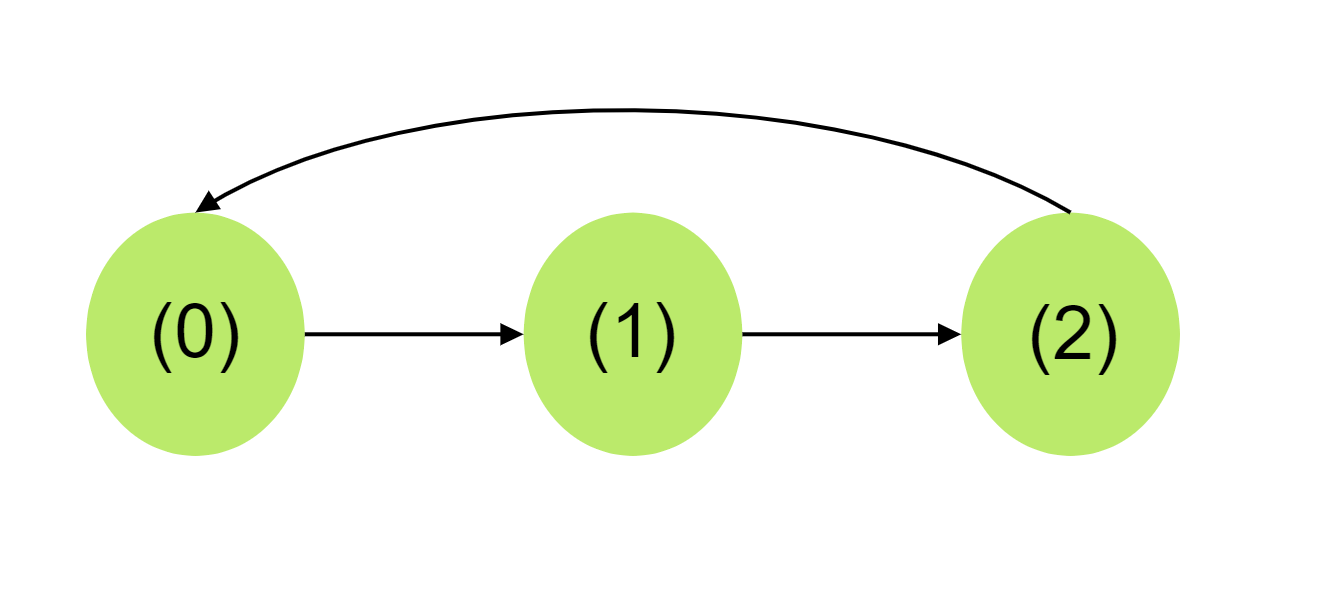}
      \caption*{Trace}
    \end{minipage}
  \end{adjustwidth}
  \caption[Label Counterexample]{Violating execution trace found by the analyzer for the label concept.}
\label{fig:label_counterexample}
\end{figure}

Outlined in listing~\ref{lst:concept_label}, the label concept describes a state machine for managing and associating distinct labels with some abstract items. It then contains functionality for querying the state for items associated with specific labels. Interestingly, while the behavior of the state machine appears to reflect the intended behavior and purpose accurately, one of the operational principles is not sound. Specifically, the Alloy Analyzer can find counterexamples for the first principle (see figure~\ref{fig:label_counterexample}) as the clear action can still happen after affixing a label. In such instances, the item will not be included in the set returned by \kw{find}. Adding \kw{clear} to the principle satisfies it. 

\begin{lstlisting}[style=concept, float=ht, caption={[Correction of Operational Principle in Label Concept] Correction of operational principle in label concept}, label={lst:concept_label_fixed}]
affix(i,l) then i in find(l) until detach(i,l) or clear(i)
\end{lstlisting}

\newpage
\subsection{Email}\label{subsec_concept_email}
The email concept defines a way for users to send content through messages. Other users can receive messages and, once in a user's inbox, they can be deleted.
\begin{lstlisting}[style=concept,float=ht,caption={[Email Concept] The email concept defined.},label={lst:conept_email}]
concept email
purpose "communicate with private messages" 
state
  inbox: User -> set Message
  from, to: Message -> User
  content: Message -> Content
actions
  send(by, for : User, m : Message, c : Content)
    when m not in for.inbox 
      m.content := c
      m.from := by
      m.to := for
  receive(by : User, m : Message)
    when m not in by.inbox and m.to = by
      by.inbox += m 
  delete(u : User, m : Message)  
    when m in u.inbox 
      u.inbox -= m
      m.from, m.to, m.content := {}
principle 
  send(f,t,m,c); receive(t,m); m in t.inbox and m.content = c
\end{lstlisting}

\section{Concept Composition}\label{sec:case_concept_composition}
Similar to section~\ref{sec:case_concepts}, this section briefly presents two examples of concept composition.

\subsection{Todo-Label App}\label{subsec:todo-label-app}
\begin{figure}[ht!]
    \centering
    \begin{minipage}{0.50\textwidth}
\begin{lstlisting}[style=concept] 
app todo_label 
include 
  todo 
  label [todo.Task, string]
sync todo.delete(t)
  label.clear(t)
sync todo.add(t)
  label.affix(t, "pending")
sync todo.complete(t)
  label.detach(t, "pending")
sync label.detach(t, "pending")
  todo.complete(t)
\end{lstlisting}
    \end{minipage}\hfill
    \begin{minipage}{0.45\textwidth}
    \includegraphics[scale=0.6]{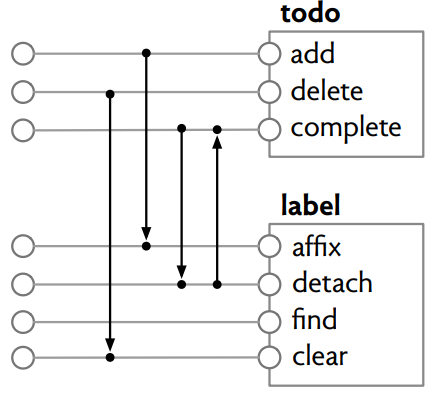}
    \end{minipage}
        \caption[Todo-Label-App]{An application synchronizing the label and todo concepts. The figure on the right is from \cite{JacksonConcepts} and depicts the synchronizations pictorially. It is worth noting that, unlike \cite{JacksonConcepts}, the \textit{Conceptual} snippet also instantiates the label type to be strings.}
        \label{fig:todo-label-app}
\end{figure}

As a first example of a composition, consider figure~\ref{fig:todo-label-app}, which synchronizes the todo and label concepts. This composition, proposed in \cite{JacksonConcepts}, attempts to represent the \textit{pending} or \textit{done} status of tasks with a built-in label \kw{"pending"}.

 A strength of this composition is that it is synergistic; it provides new functionality for querying whether tasks are pending or not using the \kw{find} action within the label concept. However, the synchronizations appear to be incomplete. Translating the model to Alloy, one can easily check the formula in listing~\ref{lst:todo-label-app-assertion} and discover counterexamples. Specifically, the label concept can still \kw{affix} the \kw{"pending"} label without the task being added to the \kw{pending} set of the todo concept. 
 
\begin{lstlisting}[style=alloy, float=ht,caption={[Violated Formula in Todo-Label-App]A formula that is true if all items with the \kw{"pending"} label are also in the \kw{pending} set of the todo concept.},label={lst:todo-label-app-assertion}]
all t : todo/Task | 
    t in label/find["pending"] =>  t in todo/State.pending
\end{lstlisting}

To satisfy the above formula, the composition must additionally include the synchronization in listing~\ref{lst:todo-label-app-sync}. 

\begin{lstlisting}[style=concept,float=ht,caption={[New Synchronization for Todo-Label-App]New synchronization for the todo-label-app to satisfy formula~\ref{lst:todo-label-app-assertion}.},label={lst:todo-label-app-sync}]
sync label.affix(t, "pending")
  todo.add(t)
\end{lstlisting}

 Ultimately, the composition would make the state component of the todo concept redundant, as the information for remembering if tasks are pending or done is now stored in the labels. Furthermore, suppose there was a richer version of the label concept that offered logical querying; it would then be possible to ask for tasks that are both \textit{pending} and \textit{urgent}. Google's Gmail 
 also synergistically uses labels for organizing sent and deleted messages, similar to what was described above.

\newpage
\begin{figure}[ht!]
    \centering
    \begin{minipage}{0.55\textwidth}
\begin{lstlisting}[style=concept]
app todo_label_mail
include
  todo [email.Content]
  label [email.Content]
  email 
sync todo.delete(t)
  label.clear(t)
sync email.receive(one todo_user, m)
  todo.add(m.content)
\end{lstlisting}
    \end{minipage}\hfill
    \begin{minipage}{0.40\textwidth}
    \includegraphics[scale=0.4]{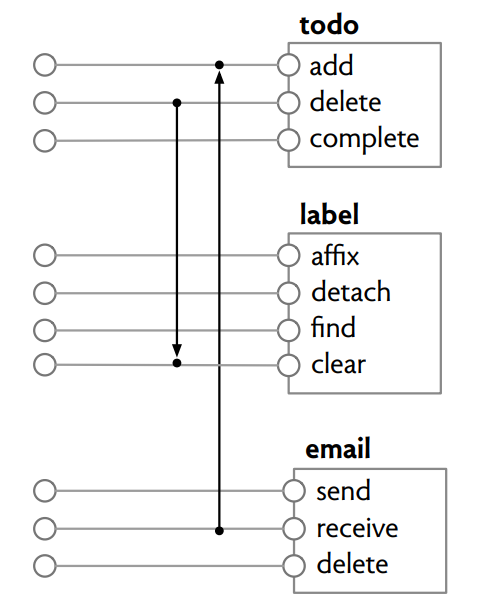}
    \end{minipage}
    \caption[Todo-Label-Email App]{A composition of todo, label, and email concepts.}
    \label{fig:todo-label-email-app}
\end{figure}

\subsection{Todo-Label-Email App}\label{case:subsec_todo_label_email_app}
\begin{figure}[ht!]
    \centering
    \includegraphics[scale=0.75]{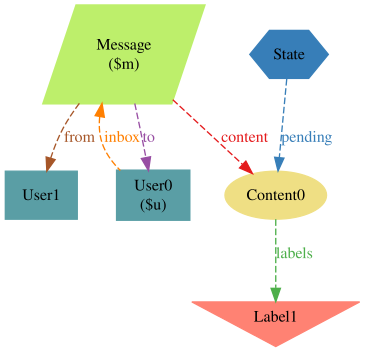}
    \caption{Possible instance from the composition in figure~\ref{fig:todo-label-email-app}.}
    \label{fig:todo_label_email_app_state}
\end{figure}

Another possible composition of the todo and label concepts includes the email concept. Figure~\ref{fig:todo-label-email-app} depicts a new synchronization that automatically creates new \textit{todos} only when a special email account \kw{todo\_user} receives emails. Although the functionality tends to be more elaborate in practice, a similar feature exists in many popular email clients or plug-ins. The Alloy analyzer can then produce a possible trace or state of the system, as seen in figure~\ref{fig:todo_label_email_app_state}.

\chapter{Concluding Remarks}\label{chap:concluding_remarks}
\textit{This chapter concludes the presented work by summarizing the main findings and outcomes of the thesis. Section~\ref{sec:conclusion_thesis_goals} assesses whether the goals presented at the beginning of the thesis have been accomplished. Then section~\ref{sec:concluding_learning_outcome} reflects on personal outcomes from doing the thesis work. Penultimately, section~\ref{sec:concluding_future_work} discusses the shortcomings of the project and potential avenues for future work. Lastly, section~\ref{conclusion:sec_final_remarks} concludes with final remarks.}

\section{Thesis Goals}\label{sec:conclusion_thesis_goals}
This section evaluates each goal introduced in section~\ref{sec:goals}, determining whether they have been achieved or not. Each goal is reiterated below for completeness, followed by a final assessment. 

\begin{enumerate}[label=\textbf{Goal \arabic*:},leftmargin=*,align=left]
    \item Develop a well-defined \gls{dsl} for concept-based design. In particular, the \gls{dsl} must define a way to represent concepts and should include language constructs for composing concepts. 
\end{enumerate}

Chapter~\ref{chap:parsing_conceptual} defines how \textit{Conceptual} parses the input, outlining how concepts and apps are reduced according to a formal, unambiguous grammar. The proposed \gls{dsl} formalizes a general syntax based on the \textit{pseudo-specifications} in \cite{JacksonConcepts}, which are partly expressed in natural language. However, certain design decisions, such as the omission of a boolean type or the distinction between querying and mutating actions, have evidently been influenced by Alloy's syntax to make the translation simple. While these decisions are argued for in a broader context, it raises the question of whether such choices have inadvertently harmed the ease of use, extensibility, or semantics of \textit{Conceptual}. Certain features, such as fresh variable declarations, have been omitted in part due to the difficulty of modeling them in Alloy. Nevertheless, the defined \gls{dsl} successfully meets all the established criteria, and the goal is considered to have been \textbf{accomplished}.

\begin{enumerate}[resume,label=\textbf{Goal \arabic*:},leftmargin=*,align=left]
 \item The \gls{dsl} should be compilable to the Alloy modeling language  \cite{Alloy6Documentation,DanielJacksonAlloy} for automated and exhaustive analysis of a specification. 
\end{enumerate}

In chapter~\ref{chap:cg_code_generation}, the thesis identifies a subset of Alloy that is effective for modeling concept specifications. Although the thesis does not include complete Alloy models due to their verbosity, it demonstrates how translated specifications can be reasoned about using Alloy's model finder. Additionally, by using the proposed compiler, a handful of errors are discovered in the specifications in \cite{JacksonConcepts}, validating the compiler's utility in enhancing the reliability of concept-based designs. Nevertheless, the compiler is only tested on a scarce number of cases. Ultimately, the goal is considered to be \textbf{accomplished}, but it lays a foundation for future testing and exploration.  

\begin{enumerate}[resume,label=\textbf{Goal \arabic*:},leftmargin=*,align=left]
    \item The syntax and semantics of the \gls{dsl} should be illustrated with several examples and reflect principles of concept-based design.  
\end{enumerate}

This goal is two-fold: it seeks to demonstrate the practicality of the \gls{dsl} through a series of examples while simultaneously ensuring that the \gls{dsl} is grounded in certain key properties of concept-based design, such as modularity and separation of concerns. Addressing the first aspect of the goal, chapter~\ref{chap:cases} showcases a variety of examples, emphasizing the \gls{dsl}'s broader applicability as opposed to analyzing the cases in-depth. Unlike \cite{JacksonConcepts}, which covers each concept in great detail, the examples were presented briefly, mostly highlighting the syntax and less so the semantics. The second part of the goal addresses more abstract concerns. Attempting to meet these standards without imposing excessive restrictions, \textit{Conceptual} omits certain general-purpose structures and implements measures at both the parsing and semantic analysis stages to enforce the independence of actions, concepts, and apps. In conclusion, ascribed partly to the absence of a more nuanced design analysis of a concrete case and the reliance on the reader's intuitive familiarity with concepts to avoid discussing their semantics, the goal is assessed to be \textbf{partially accomplished}.

\begin{enumerate}[resume,label=\textbf{Goal \arabic*:},leftmargin=*,align=left]
    \item Custom language support could be added to Visual Studio Code. 
\end{enumerate}

Although the above goal has not been discussed extensively in the main body of the thesis due to its \textit{could} prioritization, some effort has still gone into this goal. In particular, some basic initial work has been made to meet many of Visual Studio Code's language standards. As a result, the \gls{ide} now recognizes \kw{.con} files for writing \textit{Conceptual}, thereby facilitating several quality-of-life features and syntax highlighting. Nevertheless, tool support within the \gls{ide} has only been established for features that do not require an external language server, which is necessary for enabling many advanced features developers have come to expect, such as diagnostics or code completion. Thus, the goal is only considered \textbf{partially accomplished}. Moreover, the extension has only been tested locally and has not yet been made publicly available.

\section{Learning Outcome}\label{sec:concluding_learning_outcome}
Although the goals outlined in section~\ref{sec:goals} have been the focal points guiding the thesis, the ensuing process has yielded several additional personal outcomes. 

As a first outcome, the work has been paramount in fostering a deeper technical understanding of the subject matter in the author, particularly in the domains of software modeling and language design. The challenge of crafting a \gls{dsl} for modeling has also facilitated a broader exploration into other areas, such as establishing custom language support or general-purpose extensions for \glspl{ide}. Additionally, translating the \gls{dsl} to Alloy has not only imparted a thorough understanding of a completely new language but has also enhanced the author's appreciation of previously utilized modeling languages, e.g. \glsxtrshort{vdm}. While the conceivable complexity of concepts and their compositions remains uncharted territory, concept-based design seems adequate for effectively modeling many modern applications. Working with concepts has been an eye-opening experience, truly emphasizing how simple, modular, and intuitive constructs can come together to capture more complicated behavior. 

Another outcome, which is more culturally aligned, has been the opportunity to conduct and experience research first-hand at a renowned foreign university. It has been wonderful experiencing \gls{mit} and meeting so many talented and welcoming people - students, faculty, and staff alike. 
Being able to take part in various cutting-edge seminars at \glsxtrshort{csail}, featuring prominent speakers such as Prabhakar Raghavan from Google, has been delightful. Having an office in the same place that houses acclaimed researchers and groups like the Julia Lab has likewise been a surreal experience. The abundance of daily events at \gls{mit} has also contributed to the stay being more than just secluded study sessions in a different environment. It has been a wonderful and memorable time, from trivia nights with the visiting student association to a partial solar eclipse and simply experiencing sights and local cuisine in Boston and Cambridge. 

Moreover, the project work has led to an increased familiarity with academia and technical writing. The thesis is the first substantial and more profound research done independently by the author. Additionally, it is the first piece of work that has received significant and continuous feedback. In past projects, review of the written material has been completely absent during the design process. Instead, reports tended to end up as side artifacts that were magically conjured up a few days before the project submission date. Despite perhaps not being the most interesting aspect of projects, this work has shown the author the benefits of actively sharing and revising such write-ups or documentation. Uncovering errors in published, peer-reviewed material has also been an exciting first experience. 

\section{Future Work}\label{sec:concluding_future_work}
Limitations of the \gls{dsl} have generally been presented throughout the thesis in their relevant sections. Many of these limitations, though, are not attributable to some foundational technical challenges but stem from the project's constrained scope, and other features have been prioritized more. This section evaluates the shortcomings and concerns of the \gls{dsl}, proposes solutions, and advocates for future work.

Section~\ref{sec:syntax_errors} describes the handling of syntax errors in the \gls{dsl}, highlighting its susceptibility to produce poor error messages due to using a parser generator. To improve the parser's ability to provide user-friendly feedback, the \gls{dsl} should be embellished with a comprehensive set of concrete error messages in Menhir's special \kw{.messages} format \cite{MenhirReferenceManual2023}, detailing every erroneous state. These messages can then be integrated seamlessly into the compiler with minimal adjustments to the driver code. 

Another area of interest in terms of future research is related to scalability. Specifically, another mapping strategy to Alloy might be more efficient for verification than the one proposed in chapter~\ref{chap:cg_code_generation}. After deeming the translation sufficient for modeling the concepts in \cite{JacksonConcepts}, the focus in this work shifted away entirely from the quantity of constraints that needed to be analyzed by Alloy's analysis tool. The rationale behind this shift was based on the assumption that concepts are intended to be small by design. However, in hindsight, concept-based design is an entirely new modeling approach, without much empirical knowledge regarding the potential complexity of a concept or even how many concepts could feasibly be composed in practical applications. As such, to allow the analyzer to explore larger scopes, the \gls{dsl} may benefit from optimizing the number of constraints in the produced code. 

Admittedly, the current testing suite, which consists largely of the cases in chapter~\ref{chap:cases}, is not exhaustive nor provides full coverage. Certain language constructs, e.g. set comprehensions, are not tested thoroughly in the suite but have rather been manually examined in an ad-hoc manner. Moving forward, the \gls{dsl} should aim to strengthen its robustness by prioritizing establishing a more complete testing suite. Moreover, in compiler design, upon encountering bugs it is idiomatic to include them in a suite of regression tests to avoid patching the same error multiple times. The \gls{dsl} currently has no such suite for known bugs, but it is worth adding. 

Furthermore, conducting a use-case study could help validate the \gls{dsl}, particularly concerning its syntax. The process may also help unveil bugs or inconveniences and illuminate potential new features such as variable declarations, first-order quantifiers, or ways of quantifying assignments to modify subsets easily. After all, experience shows that virtually any language, whether natural or for programming, evolves with use. However, according to \cite{FowlerDSL}, it is crucial to only introduce new features in moderation, as \glspl{dsl} have a tendency to devolve into general-purpose languages. 

Another potential avenue for future work is to proceed with establishing \gls{ide} tool support. Implementing an external language server that communicates using standard protocols, e.g. \gls{lsp} \cite{microsoft_lsp_2024}, would facilitate the integration of the \gls{dsl} in any \gls{lsp}-compliant code editor. Moreover, this could introduce advanced language features, including code completion proposals, error-checking for diagnostics, semantic highlighting, information on hovering, jump-to-definitions, etc. However, this entails quite a significant amount of work. 

Finally, come a few nitpicks. Firstly, the current approach to handling generic concepts, particularly the need to change the generic types of concepts for certain synchronizations (as was the case in figure~\ref{fig:todo-label-app} or \ref{fig:todo-label-email-app}), could potentially be refined. Introducing a more agnostic type checker, perhaps similar to the Alloy Analyzer's \cite{EdwardsJonathan_Alloy_Typesystem}, may allow for greater flexibility in that one case and generally. Secondly, another simple translation may exist that does not require a top-level singleton signature for the state, which, if eliminated, could increase the readability of the generated Alloy code - especially for synchronizations where each concept has its own state component.

\section{Final Remarks}\label{conclusion:sec_final_remarks}

Modeling is fundamental for conceptualizing and analyzing systems, aiding decision-making and design optimization. However, capturing the complex behavior of a real-world system is often quite challenging while maintaining concerns such as simplicity, clarity, and reusability.

The work that has been presented in this thesis contributes to the field of modeling by formalizing a new \gls{dsl} for modeling software using concepts, which keeps the language constructs simple and highly reusable. Although concrete translations to executable code were outside the scope of this work, the expressiveness and utility of the \gls{dsl} were explored using a custom compiler to Alloy, demonstrating how behavior can be modeled using familiar concepts. Furthermore, the implemented tools were used to identify some unintended bugs in existing design specifications in the literature.

\footnotesize  %NOTE: reduced the size of the text for the bibliography
%NOTE: set the style for the bibliography and display the references used within the document
\printbibliography
\normalsize

\appendix
\chapter{List of Acronyms and Abbreviations}\label{appendix:acronyms}

Several terms and abbreviations are used throughout the thesis, many of which are technical and general computer science terms. While the thesis attempts to define acronyms the first time they are used, a few exceptions have been deliberately made for clarity's sake. This is judged case-by-case and particularly affects terms where the full name is uncommon or otherwise widely unknown. However, it is not assumed that all readers will be familiar with each term before reading this thesis. Therefore, for ease of reference and with the intent to assist readers unfamiliar with certain terms, the following comprehensive list explicitly defines each acronym used in the thesis.

{
\renewcommand{\glsnamefont}[1]{\textbf{#1}}
\printglossary[title=Acronyms \& Abbreviations,type=\acronymtype,]
}

\chapter{Lexical Tokens}\label{appendix:lexical_tokens}
As hinted at in section~\ref{sec:reading_guide}, lexical tokens are incorporated throughout the thesis using the notation: \kw{TOKEN}. However, while the context is often self-explanatory, the underlying patterns of the tokens are not explicitly mentioned anywhere. This section, and figure~\ref{fig:Token_reference} specifically, amends this vagueness by exhaustively covering each token\footnote{Except string literals (\kw{STR\_LIT}) and the special tokens such as the end-of-file marker (\kw{EOF}). String literals are handled in the standard way (see section~\ref{subsec:lexing_literals}}). Moreover, the figure demonstrates that different character sequences can generate the same token, thus enabling \textit{Conceptual}'s dual syntax. Mostly, the patterns used in figure~\ref{fig:Token_reference} are all pure string search matches. However, a few of the tokens are generated using the regular expressions depicted in listing~\ref{lst:lexing_patterns}. 

\begin{lstlisting}[float=ht,caption={[Lexing Patterns] Lexing patterns used by tokens in \textit{Conceptual}. Other shorthands include whitespace and special escape characters.},label={lst:lexing_patterns}]
let digit = ['0'-'9']
let digits = digit+
let ident_start =  ['a'-'z''A'-'Z']
let ident_continue = ['a'-'z''A'-'Z''_''0'-'9']*
let ident = ident_start ident_continue
\end{lstlisting}

For completeness, another technical detail that is not brought up succinctly in the thesis: certain tokens are parameterized. An \kw{INT\_LIT} token carries a 64-bit integer. Furthermore, \kw{STR\_LIT}, \kw{IDENT}, and \kw{ACT} each carry a string. The remaining tokens are parameterless. 

\newcommand{\keywordtoken}[2]{%
  \noindent\makebox[0.45\linewidth][l]{\texttt{#1}}%
  \makebox[0.10\linewidth][c]{$\rightarrow$}%
  \makebox[0.45\linewidth][r]{\kw{#2}}
}

\begin{figure}
    \centering
    \begin{adjustwidth}{-1.5cm}{-1.5cm}  % Adjust these values as needed for your layout
  \setlength{\columnsep}{2cm} % Set the space between columns
  {\small
    \begin{multicols}{3}
      \noindent
      \keywordtoken{+}{PLUS}
      \keywordtoken{-}{MINUS}
      \keywordtoken{\&}{AMP}
      \keywordtoken{:}{COLON}
      \keywordtoken{.}{DOT}
      \keywordtoken{,}{COMMA}
      \keywordtoken{\~}{TILDE}
      \keywordtoken{\^}{CARET}
      \keywordtoken{*}{STAR}
      \keywordtoken{/}{SLASH}
      \keywordtoken{\%}{PERCENT}
      \keywordtoken{\#}{CARD}
      \keywordtoken{(}{LPAR}
      \keywordtoken{)}{RPAR}
      \keywordtoken{[}{LBRACK}
      \keywordtoken{]}{RBRACK}
      \keywordtoken{\{}{LBRACE}
      \keywordtoken{\}}{RBRACE}
      \keywordtoken{|}{PIPE}
      \keywordtoken{<}{LT}
      \keywordtoken{>}{GT}
      \keywordtoken{<=}{LTE}
      \keywordtoken{>=}{GTE}
      \keywordtoken{=}{EQ}
      \keywordtoken{is}{EQ}
      \keywordtoken{\&\&}{LAND}
      \keywordtoken{and}{LAND}
      \keywordtoken{||}{LOR}
      \keywordtoken{or}{LOR}
      \keywordtoken{->}{ARROW}
      \keywordtoken{none}{EMPTY}
      \keywordtoken{\{\}}{EMPTY}
      \keywordtoken{empty}{EMPTY}
      \keywordtoken{when}{WHEN}
      \keywordtoken{in}{IN}
      \keywordtoken{not}{NOT}
      \keywordtoken{!}{NOT}
      \keywordtoken{set}{SET}
      \keywordtoken{one}{ONE}
      \keywordtoken{lone}{LONE}
      \keywordtoken{some}{SOME}
      \keywordtoken{const}{CONST}
      \keywordtoken{string}{STR}
      \keywordtoken{int}{INT}
      \keywordtoken{can}{CAN}
      \keywordtoken{until}{UNTIL}
      \keywordtoken{then}{THEN}
      \keywordtoken{;}{THEN}
      \keywordtoken{no}{NO}
      \keywordtoken{concept}{CONCEPT}
      \keywordtoken{purpose}{PURPOSE}
      \keywordtoken{actions}{ACTIONS}
      \keywordtoken{principle}{OP}
      \keywordtoken{app}{APP}
      \keywordtoken{include}{INCLUDE}
      \keywordtoken{sync}{SYNC}
      \keywordtoken{\textbf{ident}}{IDENT}
      \keywordtoken{ident(}{ACT}
      \keywordtoken{digits}{INT\_LIT}
      % add more as needed
    \end{multicols}
  }
\end{adjustwidth}
    \caption[Underlying Strings of Lexical Tokens]{Comprehensive overview of lexical tokens' strings and regular expressions.}
    \label{fig:Token_reference}
\end{figure}

\chapter{Complete Language Syntax}\label{appendix:complete_syntax}
This appendix provides a comprehensive overview of \textit{Conceptual}'s entire syntax, using the extended \gls{bnf} notation presented in section~\ref{sec:conceptual_parsing_ref}. Tokens are given in uppercase and are generated according to the lexical rules in appendix~\ref{appendix:lexical_tokens}.

\begin{obeliskgrammar}
\gramfunc{\primUNDERSCOREty{}}& \gramdef & \gramhor{\STR*{}, \INT*{}, \IDENT*{}} \\& & \\

\gramfunc{\mult{}}& \gramdef & \gramhor{\SET*{}, \ONE*{}, \LONE*{}}
  \\& & \\
  
\gramfunc{\ty{}}& \gramdef & 
\gramhor{\gramopt{\mult*{}} \primUNDERSCOREty*{},\primUNDERSCOREty*{} \ARROW*{} \ty*{}}
  \\& & \\

\gramfunc{\lval{}}& \gramdef & 
\gramhor{\IDENT*{},
 \lval*{} \DOT*{} \lval*{}}
  \\& & \\

\gramfunc{\call{}}& \gramdef & \ACT*{} \LPAR*{}
                               \gramseplist{\COMMA*{}}{\expr*{}} \RPAR*{}
  \\& & \\

\gramfunc{\const{}}& \gramdef & 
\gramhor{
\EMPTY{}, 
\STRUNDERSCORELIT{}, \gramopt{\MINUS{}} \INTUNDERSCORELIT{}
}
  \\& & \\

\gramfunc{\expr{}}& \gramdef & 
    \gramhor{\const*{}, 
    \unop*{} \expr*{},
    \lval*{},
    } \\
    & \grambar & \expr*{} \binop*{} \expr*{} \\
    & \grambar &\expr*{} \gramopt{\NOT*{}} \compareUNDERSCOREop*{} \expr*{}\\
   & \grambar &\expr*{} \LBRACK*{} \gramsepnelist{\COMMA*{}}{\expr*{}}
              \RBRACK*{}\\
  & \grambar &\LBRACE*{} \gramsepnelist{\COMMA*{}}{\decl*{}} \PIPE*{} \expr*{}
              \RBRACE*{}\\
  & \grambar &\gramopt*{\CAN*{} \gramopt{\NOT*{}}} \call*{}
  \\& & \\

\gramfunc{\unop}& \gramdef & 
\gramhor{
\TILDE{},
\CARET{},
\STAR{} \CARET{},
\CARD{},
\NO{}
}
  \\& & \\

    \gramfunc{\binop{}}& \gramdef & 
    \gramhor{
    \LAND{},
    \LOR{},
    \THEN{},
    \UNTIL{},
    } \\
    & \grambar &
    \gramhor{
    \PLUS{},
    \MINUS{},
    \STAR{},
    \SLASH{},
    \PERCENT{},
    } \\
    & \grambar & 
    \gramhor{
    \AMP{},
    \DOT{},
    \ARROW{},
    }
  \\& & \\

\gramfunc{\compareUNDERSCOREop{}}& \gramdef & 
\gramhor{
\EQ*{},
\IN*{},
\LT*{},
\GT*{},
\LTE*{},
\GTE*{}
}
  \\& & \\

\gramfunc{\decl{}}& \gramdef & \gramsepnelist{\COMMA*{}}{\IDENT*{}} \COLON*{}
                               \ty*{}
  \\& & \\

\gramfunc{\cUNDERSCOREsig{}}& \gramdef & \CONCEPT*{} \IDENT*{}
                                         \gramopt*{\LBRACK*{}
                                         \gramseplist{\COMMA*{}}{\IDENT*{}}
                                         \RBRACK*{}}
  \\& & \\

\gramfunc{\cUNDERSCOREpurpose{}}& \gramdef & \PURPOSE*{} \STRUNDERSCORELIT*{}
  \\& & \\

\gramfunc{\state{}}& \gramdef & \gramopt{\CONST*{}} \decl*{} \gramopt*{\EQ*{}
                                \expr*{}}
  \\& & \\

\gramfunc{\cUNDERSCOREstate{}}& \gramdef & \STATE*{} \gramstar{\state*{}}
                                           \ACTIONS*{}
  \\& & \\

\gramfunc{\assign{}}& \gramdef & 
\gramhor{
\COLON*{} \EQ*{},
\binop*{} \EQ*{}
}
  \\& & \\

\gramfunc{\stmt{}}& \gramdef & \gramsepnelist{\COMMA*{}}{\lval*{}} \assign*{}
                               \expr*{}
 \\& & \\

\gramfunc{\actionUNDERSCOREsig{}}& \gramdef & \ACT*{} \LPAR*{}
                                              \gramseplist{\COMMA*{}}{\decl*{}}
                                              \RPAR*{}
  \\& & \\

\gramfunc{\actionUNDERSCOREfiringUNDERSCOREcond{}}& \gramdef & \gramopt*{\WHEN*{}
                                                               \expr*{}}
  \\& & \\

\gramfunc{\action{}}& \gramdef & \actionUNDERSCOREsig*{}
                                 \actionUNDERSCOREfiringUNDERSCOREcond*{}
                                 \gramstar{\stmt*{}}\\
  & \grambar &\actionUNDERSCOREsig*{} \COLON*{} \ty*{} \expr*{}
  \\& & \\

\gramfunc{\cUNDERSCOREactions{}}& \gramdef & \ACTIONS*{}
                                             \gramplus{\action*{}} \OP*{}
  \\& & \\

\gramfunc{\cUNDERSCOREop{}}& \gramdef & \OP*{}
                                        \gramseplist{\COMMA*{}}{\expr*{}}
  \\& & \\

\gramfunc{\filename{}}& \gramdef & \IDENT*{}\\
  & \grambar &\IDENT*{} \DOT*{} \IDENT*{}
  \\& & \\

\gramfunc{\pathUNDERSCOREexpr{}}& \gramdef & \pathUNDERSCOREexpr*{} \DOT*{}
                                             \DOT*{} \SLASH*{}\\
  & \grambar &\pathUNDERSCOREexpr*{} \IDENT*{} \SLASH*{}\\
  & \grambar &\grameps
  \\& & \\

\gramfunc{\filepath{}}& \gramdef & \pathUNDERSCOREexpr*{} \filename*{}
  \\& & \\

\gramfunc{\appUNDERSCOREdep{}}& \gramdef & \filepath*{} \gramopt*{\LBRACK*{}
                                           \RBRACK*{}}\\
  & \grambar &\filepath*{} \LBRACK*{}
              \gramsepnelist*{\COMMA*{}}{\gramopt*{\IDENT*{} \DOT*{}}
              \primUNDERSCOREty*{}} \RBRACK*{}
  \\& & \\

\gramfunc{\syncUNDERSCOREarg{}}& \gramdef & \gramopt{\mult*{}} \expr*{}
  \\& & \\

\gramfunc{\multUNDERSCOREcall{}}& \gramdef & \ACT*{} \LPAR*{}
                                             \gramseplist{\COMMA*{}}{\syncUNDERSCOREarg*{}}
                                             \RPAR*{}
  \\& & \\

\gramfunc{\trigger{}}& \gramdef & \IDENT*{} \DOT*{} \multUNDERSCOREcall*{}
  \\& & \\

\gramfunc{\response{}}& \gramdef & \IDENT*{} \DOT*{} \call*{}
  \\& & \\

\gramfunc{\sync{}}& \gramdef & \sync*{} \trigger*{} \gramplus{\response*{}}
  \\& & \\

\gramfunc{\app{}}& \gramdef & \APP*{} \IDENT*{} \INCLUDE*{}
                              \gramplus{\appUNDERSCOREdep*{}}
                              \gramstar{\sync*{}}
  \\& & \\

\gramfunc{\concept{}}& \gramdef & \cUNDERSCOREsig*{} \cUNDERSCOREpurpose*{}
                                  \cUNDERSCOREstate*{} \cUNDERSCOREactions*{}
                                  \cUNDERSCOREop*{}
  \\& & \\

\gramfunc{\program{}}& \gramdef & \gramstar{\concept*{}} \gramstar{\app*{}}
                                  \EOF*{}
  \\& & \\

\end{obeliskgrammar}
\label{appendix:language_support_vscode}

\chapter{Language Support in Visual Studio Code}
This appendix details the extent to which the \textit{Conceptual} is supported in the Visual Studio Code \gls{ide}. Currently, the utilities discussed in this appendix are exclusively available for local use and have not been released to the broader extension ecosystem. This appendix will not discuss the compiler pipeline proposed in chapter~\ref{chap:compiler}, which, likewise, is yet to be made publically available in the \gls{ide}.

\section{General Language Utilities}
The \gls{dsl} is supported at a basic level in Visual Studio Code. The code editor now recognizes \kw{.con} files for writing \textit{Conceptual}, supporting several quality-of-life features, including auto-indentation, code folding, single- and multi-line commenting, GitHub Copilot integration, as well as automatic closure of sets of parentheses, brackets, braces, quotation marks, and so forth. The language is configured in a common, standardized way to enable developers to use their existing \gls{ide}-settings and shortcuts 

Support has been implemented only for features that do not necessitate an external language server, meaning functionalities such as code completion, semantic highlighting, and jump-to-definitions are not supported. Consequently, the \gls{ide}-support that is currently established runs quite efficiently as language servers add a significant amount of overhead, especially during the initial loading and analysis of a project. The advantages of including a language server, though, tend to outweigh the costs. 

\section{Syntax Highlighting}
Syntax highlighting is a key feature of modern code editors that enhances readability and error detection in programs, making coding more efficient. By providing a special \textit{TextMate} grammar, which is a structured collection of Oniguruma\footnote{Popular library for regular expressions that supports several character encodings.} regular expressions written in \glsxtrshort{xml} or \glsxtrshort{json}, syntax highlighting can be obtained leveraging Visual Studio Code's tokenization engine. This engine runs in the same process as the renderer, and tokens are updated as they are typed. 

Furthermore, each token is associated with a particular scope that defines its context. To broadly support grammar, \textit{Conceptual} has been written according to common conventions, largely building on existing scopes that many themes target rather than defining new ones. Consequently, \kw{.con} files are directly compatible with several existing themes and styles within the \gls{ide}.  

\begin{figure}[ht!]
    \centering
    \includegraphics[width=0.9\linewidth]{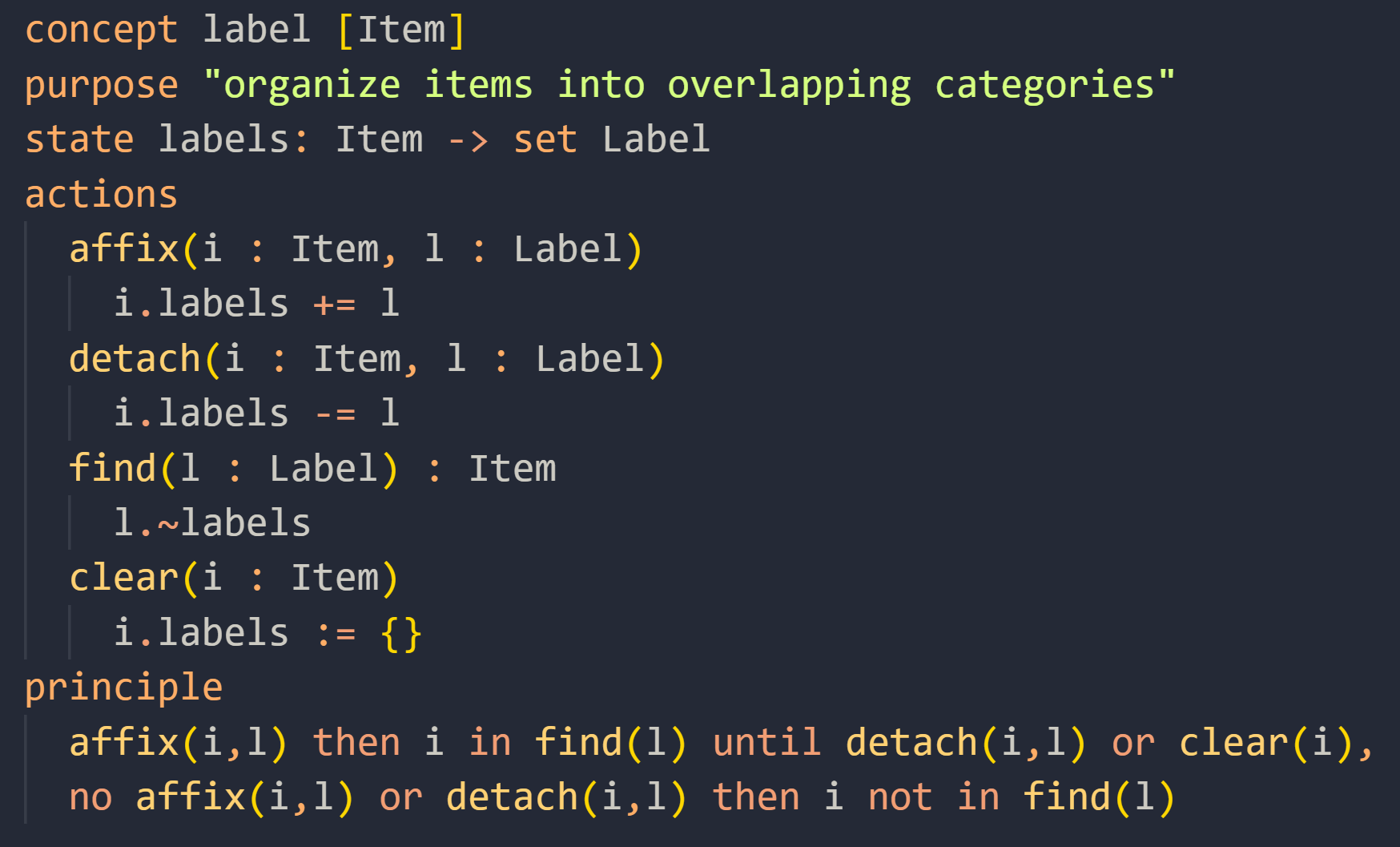}
    \caption[Example of Syntax Highlighting]{Example of syntax highlighting using the \textit{Ayu Mirage Bordered} theme.}
    \label{fig:example-syntax-highlighting}
\end{figure}

\end{document}